\documentclass[envcountsame,runningheads]{llncs}
\usepackage{pscproc2}
\usepackage{fixltx2e}

\usepackage{xspace,amsmath}
\usepackage{url}

\usepackage{color}
\usepackage{booktabs}

\usepackage{hyperref}

\usepackage{graphicx}
\usepackage{epstopdf}
\usepackage{epsfig}

\newcommand{\exclude}[1]{}

\definecolor{red}{rgb}{1.0,0,0}

\DeclareMathOperator{\argmin}{argmin} 

\begin{document}

\title{Faster range minimum queries}

\author{Tomasz Kowalski, Szymon Grabowski}
\institute{$^\dag$ Lodz University of Technology, 
           Institute of Applied Computer Science,\\
           Al.\ Politechniki 11, 90--924 {\L}\'od\'z, Poland, \\
           \email{\{tkowals|sgrabow\}@kis.p.lodz.pl}
}

\maketitle

\begin{abstract}
Range Minimum Query (RMQ) is an important building brick of many 
compressed data structures and string matching algorithms.
Although this problem is essentially solved in theory,
with sophisticated data structures allowing for constant time queries,
practical performance and construction time also matter.
Additionally, there are offline scenarios in which the number of queries, 
$q$, is rather small and given beforehand, 
which encourages to use a simpler approach.
In this work, we present a simple data structure, with very fast construction, 
which allows to handle queries in constant time on average.
This algorithm, however, requires access to the input data during queries 
(which is not the case of sophisticated RMQ solutions).
We subsequently refine our technique, combining it with 
one of the existing succinct solutions with $O(1)$ worst-case 
time queries and no access to the input array.
The resulting hybrid is still a memory frugal data structure,
spending usually up to about $3n$ bits, and providing competitive query times, 
especially for wide ranges.
We also show how to make our baseline data structure more compact.
Experimental results demonstrate that the proposed BbST (Block-based Sparse Table) 
variants are competitive to existing solutions, also in the offline scenario.
\end{abstract}

\begin{keywords}
string algorithms, range minimum query, bulk queries
\end{keywords}

\section{Introduction}

The Range Minimum Query (RMQ) problem is to preprocess an array 
in a way allowing to return 
the position of the minimum element for an arbitrary input interval, 
specified by a pair of indices, 
in an efficient manner.
More formally, for an array $A[1 \ldots n]$ of objects from a totally ordered universe 
and two indices $i$ and $j$ such that $1 \leq i \leq j \leq n$, 
the range minimum query $\mathsf{RMQ_A}(i, j)$ returns $\argmin_{i\leq k\leq j} A[k]$, 
which is the position of a minimum element in $A[i \ldots j]$.
One may alternatively require the position of the leftmost minimum element, 
i.e., resolve ties in favour of the leftmost such element, 
but this version of the problem is not widely accepted.
In the following considerations we will assume that $A$ contains integers 
from the universe $U = \{1, 2, \ldots, n\}$, of $\log_2 n$ bits each.

This innocent-looking little problem has quite a rich and vivid history and perhaps 
even more important applications, in compressed data structures in general, 
and in text processing in particular.
Solutions for RMQ which are efficient in both query time and 
preprocessing space and time are building blocks in such succinct data structures as, 
e.g., suffix trees, two-dimensional grids or ordinal trees. 
They have applications in string mining, document retrieval, bioinformatics, 
Lempel-Ziv parsing, etc. 
For references to these applications, see~\cite{FH11,FN16}.

The RMQ problem history is related to the LCA (lowest common ancestor) 
problem defined for ordinal trees: 
given nodes $u$ and $v$, return $LCA(u, v)$, which is the lowest node
being an ancestor of both $u$ and $v$. 
Actually, the RMQ problem is linearly equivalent to the LCA problem~\cite{GabowBT84,BFC2000}, 
by which we mean that both problems 
can be transformed into each other in time linearly proportional 
to the size of the input.
It is relatively easy to notice that 
if the depths of all nodes of tree $T$ visited during an Euler tour 
over the tree are written to array $A$, 
then finding the LCA of nodes $u$ and $v$ is equivalent to finding 
the minimum in the range of $A$ spanned between the first visits to $u$ 
and $v$ during the Euler tour (cf.~\cite[Observation~4]{BFC2000}).
Harel and Tarjan~\cite{HT84} were the first to give $O(n)$-time tree 
preprocessing allowing to answer LCA queries in constant time.
The preprocessing required $O(n)$ words of space.
Bender and Farach~\cite{BFC2000} presented a significantly simpler algorithm 
with the same time and space complexity.
Further efforts were focused on reducing the space of the LCA/RMQ solution, 
e.g., Sadakane~\cite{Sadakane07} showed that LCAs on a tree of $n$ nodes 
can be handled in constant time using only $2n + o(n)$ bits.
A crowning achievement in this line of research was the algorithm of 
Fischer and Heun~\cite{FH11}, who showed that RMQs on $A$ can be transformed 
into LCA queries on the succinct tree, and this leads to an RMQ solution 
that also uses $2n + o(n)$ bits and (interestingly) does not access $A$
at query time.
This result essentially matches the information-theoretic lower bound 
for an RMQ solution not accessing the input array, which is $2n - \Theta(\log n)$ bits.
Any scheme for RMQs allows to reconstruct the Cartesian tree~\cite[Section~2.2]{FH11} 
of the input array by iteratively querying the scheme for the minimum; 
the number of bits to describe any possible Cartesian tree here is 
$2n - \Theta(\log n)$~\cite{Jacobson89,FH11}, hence the bound.

The Fischer and Heun solution, although allowing for constant time RMQ queries, 
is not so efficient in practice: handling one query takes several 
microseconds (see~\cite{FN16}).
Some ingenious algorithmic engineering techniques, 
by Grossi and Ottaviano~\cite{GrossiO13}, Ferrada and Navarro~\cite{FN16}, 
and Baumstark et al.~\cite{BaumstarkGHL17}, 
were proposed to reduce this time, 
and the fastest implementation~\cite{BaumstarkGHL17} achieves around 
$1\mu$s per query (timings vary depending on query parameters) 
on an single core of the Intel Xeon E5-4640 CPU.

Recently, Alzamel et al.~\cite{ACIP17} (implicitly) posed 
an interesting question:
why should we use any of these sophisticated data structures for RMQ
when the number of queries is relatively small and building the index 
(even in linear time, but with a large constant) and answering then 
the queries (even in constant time each, but again with a large constant) 
may not amortize?
A separate, but also important point is that if we can replace a heavy tool 
with a simpler substitute (even if of limited applicability), 
new ideas may percolate from academia to software industry.
Of course, if the queries $[\ell_i, r_i]$ are given one by one, 
we cannot answer them faster than in the trivial 
$O(r_i - \ell_i + 1) = O(n)$ time for each, 
but the problem becomes interesting if they are known beforehand.
The scenario is thus offline 
(we can also speak about {\em batched queries} or {\em bulk queries}).
Batched range minima (and batched LCA queries) have applications 
in string mining~\cite{FMV08}, text indexing and various non-standard pattern 
matching problems, for details see~\cite[Section~5]{ACIP17}.

In this paper we first present a heuristical idea for RMQ computation
(without a constant-time guarantee).
This idea is very simple, the corresponding data structure very fast to build 
(as opposed to any other RMQ algorithm we are aware of) 
and it answers range minimum queries faster on average than
competitive algorithms, except perhaps on narrow intervals.
Then, a hybrid of our solution with most efficient constant-time RMQs is presented,
with usually less than $3n$ 
bits of space and no need to access $A$.
In this way we boost the average performance of constant-time solutions without
sacrificing much in the space usage.
Ideas for making our data structure compact are discussed in a separate section.
We also discuss the scenario of running batched range minima (relevant when
the number of queries is significantly smaller than the input array size),
to which we also adapt our idea.


The roadmap of the paper is as follows.
In the next section we present our block-based approach to the standard (online) 
RMQ problem.
By ``online RMQ'' we mean the scenario in which the data structure is built first, 
to handle any number of queries to follow.
In several subsections, we present a plain block-based sparse table (BbST) algorithm, 
a hybrid with a theoretical solution, a two-level BbST representation, 
and a compact representation.
Section 3 deals with the offline RMQ problem variant. 
Here the number of input queries is expected to be small compared to the input array size, 
and for this reason building a costly data structure may be an overkill. 
In this scenario we measure (both in complexity terms and in experiments) 
the time and space to handle $q$ queries.
The subsections of Section 3 present the first (and only so far) algorithm 
for offline RMQ, by Alzamel et al., and our adaptation of BbST to this setting.
Section 4 contains experimental results. 
The last section concludes.

We use a standard notation in the paper.
All logarithms are of base 2.
The space usage is sometimes expressed in words (of $\log_2 n$ bits), 
sometimes in bits, whichever more convenient, 
and we are explicit about those units.

A (very) preliminary version of our paper was presented in Proc. PSC 2017~\cite{GK17}.

\section{Our algorithms}
\label{sec:our}

\begin{figure}[pt!]
\centerline{
\includegraphics[width=0.99\textwidth,scale=1.0]{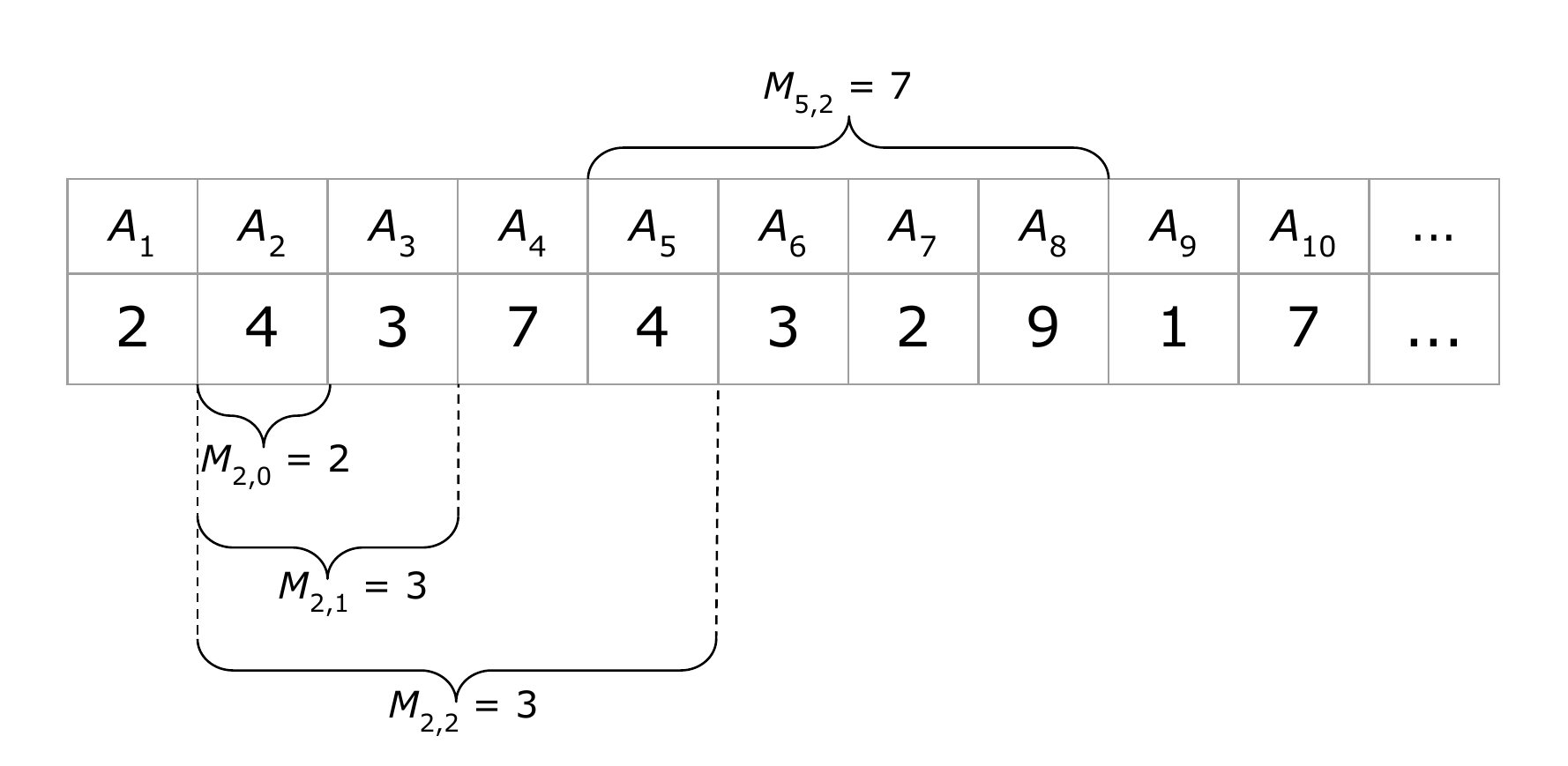}
}
\caption[ST]
{A prefix of the input array $A$, together with several precomputed Sparse Table values.
Each $M_{i,j}$ stores the position of the minimum in $A[i \ldots i + 2^j - 1]$.
For example, $M_{5,2} = 7$, since the minimum of $A[5 \ldots 8]$, which is 2, 
is located in $A[7]$ (ties are resolved arbitrarily).}
\label{fig:st}
\end{figure}

Before presenting our algorithms, we must remind the 
classic idea of the Sparse Table (ST)~\cite{BFC2000}, as a point of departure.
Given an array $A$ of size $n$, 
we compute and store the minima for all its subarrays of size being a power of two.
Let $M_{i,j}$ denote the position of the minimum of the subarray $A[i \ldots i + 2^j - 1]$
(observe that for $j = 0$ we have a subarray with one element).
Figure~\ref{fig:st} illustrates.
Any interval in $A$, with its boundaries denoted by 
$[\ell, r]$,
can be covered by a pair of such subarrays, 
for example $A[2 \ldots 8]$ is covered with $A[2 \ldots 5]$ and $A[5 \ldots 8]$. 
Finding the position of the minimum in this interval, that is, 
returning $\mathsf{RMQ_A}(\ell, r)$, 
boils down to reading two precomputed minima and 
returning the position of the minimum of these two values.
In our example, $\min(A[2 \ldots 8])$ is equal to $M_{2,2}$ 
if $A[M_{2,2}] \leq A[M_{5,2}]$, or equal to $M_{5,2}$ otherwise.
The space used by ST is $O(n\log n)$ words.

In the following subsections,
first we present our block-based sparse table idea, which competes practically 
(although not in the worst case) with existing RMQ algorithms.
This algorithm, however, requires access to $A$.
Then we propose a hybrid algorithm, improving the worst-case time and also 
allowing to get rid of $A$.

\subsection{Block-based Sparse Table}
\label{sec:our1}
The first algorithm we present can be considered as a generalization 
of ST for blocks (with some twist).

In the construction, array $A$ is divided into equal blocks of size $k$ 
and for each block $B_i$ (where $i = 1, 2, \ldots$) we find and store the 
positions of $O(\log (n/k))$ minima, where $j$th value ($j = 1, 2, \dots$) 
is the minimum of $A[(i-1)k + 1 \ldots (i-1)k + 2^{j-1} k]$, 
i.e., the minimum over a span of $2^{j-1}$ blocks, where the leftmost 
block is $B_i$.
The required space is $O((n/k)\log(n/k))$ words.

To answer a query $[\ell, r]$, 
we find 
$m'$, the position of the minimum over the
{\em smallest} span of blocks {\em fully including the query}\footnote{In the 
earlier (conference) version of our work~\cite{GK17} this algorithm was slightly 
different: 
we started from the {\em largest} span of blocks {\em fully included in the query}.
Although the space usages and time complexities of both variants are the same, 
the new one is practically faster by about 
20 percent for large intervals, 
while the speed for small intervals is more or less the same.}, 
using the technique of the sparse table.
If $\ell \leq m' \leq r$, then $m'$ is the answer and it is returned.
In the opposite (rare) case, we continue with finding 
the position $m''$ of the minimum over 
the {\em largest} span of blocks {\em fully included in the query}, 
and then scan (at most) $O(k)$ cells of $A$ to find the true minimum
and return its position.

\begin{figure}[pt!]
\centerline{
\includegraphics[width=0.99\textwidth,scale=1.0]{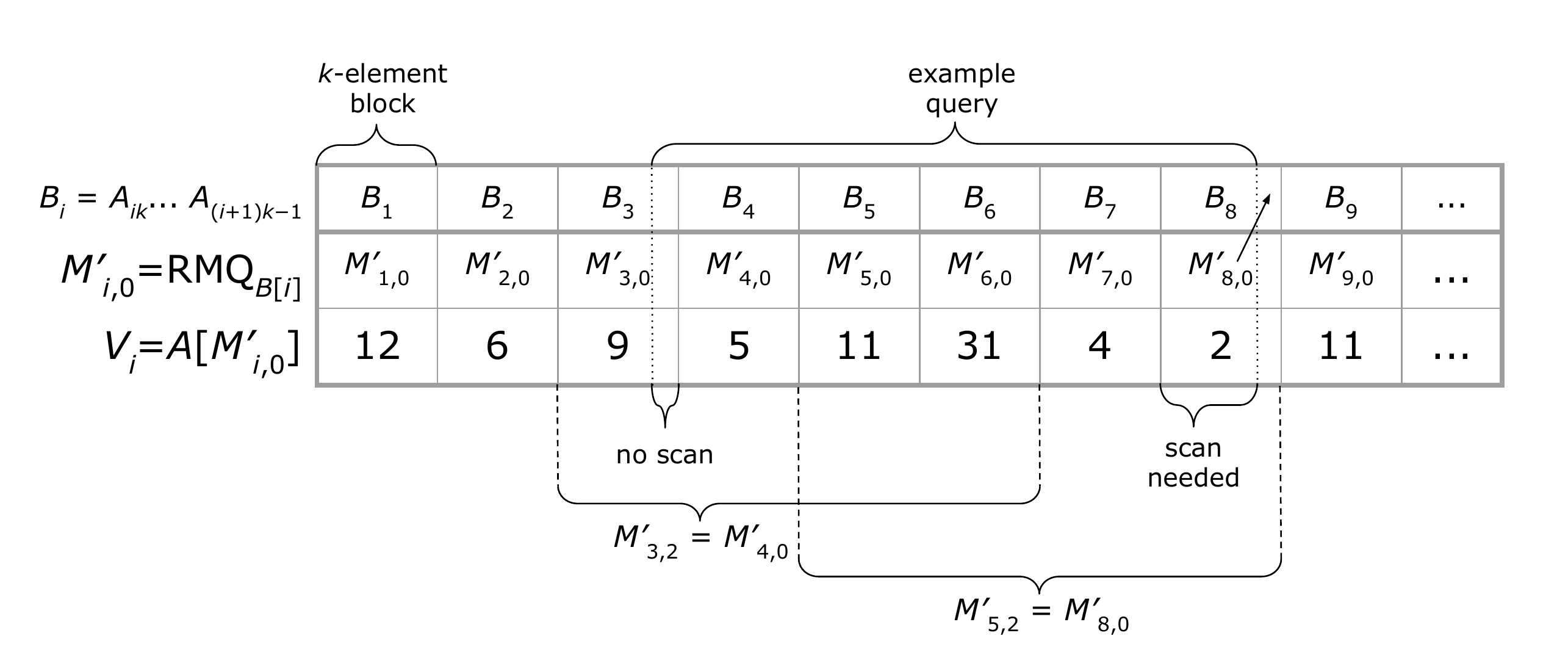}
}
\caption[FigBbST]
{A prefix of the input array $A$ divided into blocks $B_i$.
The positions of the block minima are stored in $M'_{i,0}$.
The corresponding minima values are in $V_i$.}
\label{fig:bbst}
\end{figure}

Figure~\ref{fig:bbst} reveals more details concerning this operation.
The query starts somewhere in block $B_3$ and ends in block $B_8$.
The left of the two intervals covering the range of blocks $B_3 \ldots B_8$ 
spans over $B_3 \ldots B_6$.
Each $M'_{i,j}$ ($j \geq 0$) stores the position of the minimum of the subarray of $A$ 
covering the blocks $B_i \ldots B_{i + 2^j - 1}$.
In our example, the position of the minimum for the range of $B_3 \ldots B_6$ 
is stored in $M'_{3,2}$.
The value of this minimum is 5 and it belongs to $B_4$ 
(which is denoted as $M'_{5,2} = M'_{8,0}$).
As block $B_4$ is wholly contained in our query, there is no need to scan any part 
of a block.
The situation is different for the right of the two intervals covering $B_3 \ldots B_8$, 
namely $B_5 \ldots B_8$.
Here the minimum belongs to $B_8$ (i.e., $M'_{5,2} = M'_{8,0}$) and moreover, 
it happens that it is located beyond the prefix of this block covered by the query, 
which is shown in the figure with an arrow.
In this case, the prefix of $B_8$ must be scanned.
At the end, one extra comparison returns the RMQ value.

If the average query interval width is $u$, the probability that $m'$ 
belongs to the union of the two blocks containing $\ell$ and $r$, respectively, 
is $O(k/u)$.
The average query time complexity is thus 
$O((k/u) \times k + (1 - k/u) \times 1) = O(k^2/u)$, 
which is constant for $u = \Omega(k^2)$.

{\em Sparse Table with higher arity.}
Let us now consider a generalization of the doubling technique in Sparse Table 
(a variant that we have not implemented).
Instead of using powers of 2 in the 
formula $A[(i-1)k + 1 \ldots (i-1)k + 2^{j-1} k]$, 
we use powers of an arbitrary integer $\ell \geq 2$
(in a real implementation it is convenient to assume that $\ell$ 
is a power of 2, e.g., $\ell = 16$).
Then, the minimum over a range will be calculated as a minimum 
over $\ell$ precomputed values.
The worst-case query time becomes $O(\ell + k)$, but the space 
gets reduced by a factor of $\log\ell$.
We will come back to this variant in Section~\ref{sec:offline_our}.

\subsection{A hybrid algorithm}
\label{sec:our2}

The algorithm presented in the previous subsection has two drawbacks.
One is the worst-case query time of $O(k)$, rather than $O(1)$.
The other is that it requires access to array $A$, which typically occupies 
around $n\log n$ bits.
We present now a hybrid of our technique with any existing algorithm 
with no access to $A$ at query time and constant time queries.
Such solutions, e.g., Baumstark et al.~\cite{BaumstarkGHL17}, 
may use $2n + o(n)$ bits of space. 

The hybrid builds both 
a data structure from Baumstark et al. 
and a block-based sparse table, storing however both the minimum positions 
and their values in the latter component.
Note that the plain BbST does not store the minimum values 
since $A$ is available there.

The queries in our hybrid are handled as follows. 
First the BbST component tries to answer the query. 
Using the sparse table requires comparing values of two retrieved minima 
and this is when we would have to refer to $A$, but in the modified 
BbST we access the stored values (there are only $O((n/k) \log(n/k))$ of them 
in total, not $O(n\log n)$).
If, however, we are unlucky and in the plain BbST we would have to scan 
at most two blocks, we switch to the component with $O(1)$ time.
To sum up, our solution speeds up the queries performed by Baumstark et al. 
in many practical cases, 
preserves the constant worst case time and increases the space usage only 
moderately (to less than $3n$, as we will see in the experimental section). 
The last property, compact space, requires however an appropriate representation 
of the BbST component (and well chosen parameters), 
which is described in 
Section~\ref{sec:our3}.

\subsection{Two-level block-based Sparse Table}
\label{sec:our4}

We come back to our basic variant, from Section~\ref{sec:our1}, 
and show how to generalize this procedure 
to two levels of blocks.

The idea is to compute minima for $n/k_2$ non-overlapping blocks of size $k_2$ 
and then apply the doubling technique from Sparse Table on larger blocks, 
of size $k_1$.
We assume that $k_2$ divides $k_1$.

The first construction stage, finding the minima for blocks of size $k_2$, 
takes $O(n)$ time.
The second stage, working on blocks of size $k_1$, 
takes $O(n/k_2 + (n/k_1)\log(n/k_1))$ time.
Then we answer the queries;
if we are unlucky and 
one or two blocks of size $k_1$ have to be scanned, the procedure is sped up 
with aid of the precomputed minima for the blocks of size $k_2$.
Here were assume that the queries are sampled uniformly randomly over 
the whole input array, i.e., the average query width is $O(n)$.
A query is thus answered in $O(k_1/k_2 + k_2)$ time in the worst case 
and in $O(1)$ time on average if $(k_1 / n) \times (k_1/k_2 + k_2) = O(1)$.
The condition on the average case becomes clear when we notice that 
the probability of the unlucky case is, 
under the given assumption, $\Theta(k_1/n)$ 
and checking (up to) two blocks takes $O(k_1 / k_2 + k_2)$ time.
Fulfilling the given condition implies that 
$k_1 k_2 = O(n)$ and $k_1 / k_2 = O(n/k_1)$.

Our goal is to find such $k_1$ and $k_2$ that the extra space is minimized 
but the average constant time preserved.
To this end, we set $k_1 = \sqrt{n \log n}$, $k_2 = \sqrt{n / \log n}$,
and for these values the average time becomes 
$O(1)$.
The space is $O(n/k_2 + (n/k_1)\log(n/k_1)) = 
O(\sqrt{n \log n})$ words.

Note that we preserved the average time of the variant 
from Section~\ref{sec:our1} and reduced the extra space 
by a factor of $\log^{1/2}n$.
Note also that the space complexity cannot be reduced for any 
other pair of $k_1$ and $k_2$ such that
$k_1 k_2 = O(n)$.

It is quite easy to notice that generalizing the presented scheme to 
multiple levels does not help, i.e., it is impossible to obtain both $O(1)$ 
average query time and $o(\sqrt{n\log n})$ words of space.
Indeed, let us have $h \geq 2$ levels 
and choose the parameters $k_1 > \ldots > k_h$, 
such that each $k_{i+1}$ divides $k_i$.
The minima for non-overlapping blocks of size $k_i$, 
$i = h, h-1, \ldots, 2$, are first computed, 
and then also the minima for blocks of size $k_1$, 
their doubles, quadruples, and so on.
The constant average time for query answering 
now requires that 
$(k_1 / n) \times (k_1 / k_2 + k_2 / k_3 + \ldots + k_{h-1} / k_h + k_h) = O(1)$.
The second factor on the left-hand side is $\Omega(k_h)$, 
hence the condition implies that $k_1 k_h = O(n)$ 
(which is analogous to the condition required for the two-level variant).
As the space is 
$\Theta((n/k_1)\log(n/k_1) + n/k_2 + n/k_3 + \ldots + n/k_h) = 
\Omega((n/k_1)\log(n/k_1) + n/k_h)$, it is minimized for 
$k_1 = \Theta(\sqrt{n \log n})$ and $k_2 = \Theta(\sqrt{n / \log n})$,
which gives $\Omega(\sqrt{n \log n})$ words of space, not better than 
for the case of $h = 2$.

We implemented the two-level variant, as will be seen in the experimental section.
In the standard (non-compact) version we have $k_2 \leq 256$ 
and thus the respective minimum positions are stored on one byte each.

\subsection{Compacting BbST}
\label{sec:our3}

In the (block-based) sparse array in each block we store multiple minimum positions: 
for a span of one block, a span of two blocks, a span of four blocks, and so on.
Let us denote the (conceptual) array containing the minimum positions for all spans 
over $2^j$ blocks with the {\em $j$th layer}, where $0 \leq j \leq \lfloor \log n \rfloor$.

\begin{figure}[pt!]
\centerline{
\includegraphics[width=0.99\textwidth,scale=1.0]{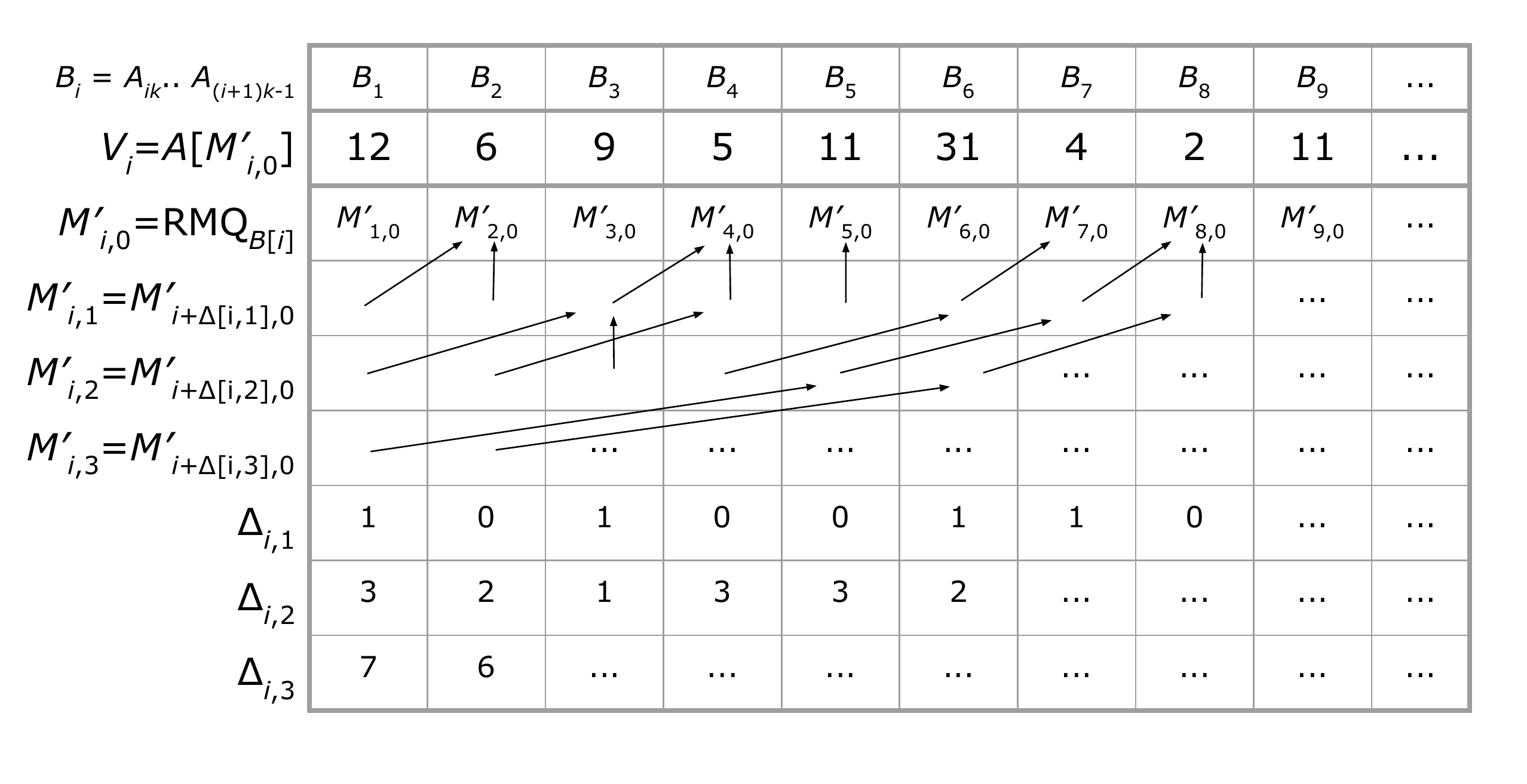}
}
\caption[FigcBbST]
{Compact variant of the block-based sparse table. 
Apart from the minimum positions for single blocks (array $V$) 
also $\Delta_{i,*}$ arrays are stored, 
which allow to access the minima over larger spans of blocks.}
\label{fig:cbbst}
\end{figure}

If we store the minimum positions na{\"i}vely in $\log n$ bits, 
the total size of our data structure is $O((n/k) \log(n/k) \log n)$ bits.
Pointing to a minimum in $j$th layer, however, requires fewer bits: $\log k$ bits 
in 0th layer and $j$ (extra) bits in $j$th layer for $j > 0$.
We simply point to a block containing a minimum rather than to
its exact position, except for the lowest layer.
Figure~\ref{fig:cbbst} presents the relation between 
$M'_{i,j}$, the position of the minimum of the span of blocks $B_i \ldots B_{i + 2^j - 1}$, 
and $\Delta_{i,j}$ which stores $i$th value for $j$th layer.
The relation is simple: $M'_{i,j} = M'_{i + \Delta_{i,j},0}$.
In this way, we can reduce the overall space to $O((n/k) (\log k + \log n) + (n/k) \log^{2}(n/k))$ 
bits.
In the real implementation, however, we store the minimum positions every $9$th layer 
directly (using $\log n$ bits) and in the remaining layers use 8 bits, i.e., 1 byte 
for a reference.
This is a convenient tradeoff between memory use and byte-aligned access to data.

We can do better though, for the price of more costly access to the minima.
To this end, each $\Delta_{i,j}$, $j > 0$, can be encoded on one bit, with reference 
to $\Delta_{i,j-1}$, 
and the total space use is then $O((n/k) (\log k + \log(n/k))$ bits.

We admit that our data structure resembles the tournament tree 
and its compact version, the navigation pile~\cite{katajainen2003navigation}.
The tournament tree is form of a min-heap, in which every leaf 
represents a player and every internal node stores a copy 
of the winner.
In the navigation pile there is no redundancy, only single bits 
telling the winner, with a primary motivation to reduce cache misses, 
e.g., in priority queue operations.

There is one more aspect concerning the required space.
Let us consider a hybrid of an $O(1)$-time RMQ algorithm with our two-level BbST 
variant (which, as experiments will show, is an attractive solution).
Apart from the minimum positions for each block of size $k_2$ we also need 
to store its value. 
In order not to spend $\log n$ bits for such values, 
we apply a quantization heuristic.
The smallest and the largest minimum among the minima for blocks of size $k_2$ 
are converted to $0$ and $maxQ$, respectively.
The other minimum values are quantized more `densely' for smaller and 
less densely for larger values (as, assuming a uniformly random distribution of 
the input data in $A$, minima for blocks tend to be closer to the global minimum 
than to the maximum among those minima for blocks).

\section{Offline Range Minimum Queries}
\label{sec:offline}

If the queries to the array are known beforehand and their number $q$ is limited, 
resigning from heavy RMQ machinery in favour of much simpler solutions 
is not only more natural, but may also prove faster and memory frugal.
In the first subsection below, we present the only known so far solution 
to this scenario, from Alzamel et al.~\cite{ACIP17}, 
while in the next subsections we show how to adapt our block-based sparse table 
to offline queries.

\subsection{The Alzamel et al. algorithm}
\label{sec:offline_Alzamel}

Following~\cite{AS14} (see the proof of Lemma 2), the Alzamel et al. approach 
starts from contracting the array $A$ into $O(q)$ entries.
The key observation is that if no query starts or ends with an index $i$ and $i+1$, 
then, if $A[i] \neq A[i+1]$, $\max(A[i], A[i+1])$ will not be the answer to 
any of the queries from the batch.
This can be generalized into continuous regions of $A$.
Alzamel et al. mark the elements of $A$ which are either a left or a right endpoint
of any query and create a new array $A_Q$: 
for each marked position in $A$ its original value is copied into $A_Q$,
while each maximal block in $A$ that does not contain a marked position is replaced 
by a single entry, its minimum. 
The relative order of the elements copied from $A$ is preserved in $A_Q$, 
that is, in $A_Q$ the marked elements are interweaved with representatives of 
non-marked regions between them.
As each of $q$ queries is a pair of endpoints, $A_Q$ contains up to $4q + 1$ elements 
(repeating endpoint positions imply a smaller size of $A_Q$, but for relative small 
batches of random queries this effect is rather negligible).
In an auxiliary array the function mapping from the indices of $A_Q$ into the original 
positions in $A$ is also kept.

For the contracted data, three procedures are proposed. 
Two of them, one offline and one online, are based on existing RMQ/LCA algorithms 
with linear preprocessing costs and constant time queries. 
Their practical performance is not competitive though.
The more interesting variant, \textsf{ST-RMQ\textsubscript{CON}}, 
achieves $O(n + q\log q)$ time\footnote{Written consistently as $n + O(q\log q)$ 
in the cited work, to stress that the constant associated with scanning the original 
array $A$ is low.}.
The required space (for all variants), on top of the input array $A$ and the list of 
queries $Q$, is claimed to be $O(q)$, but a more careful look into the algorithm 
(and the published code) reveals that in the implementation of the contracting step 
the top bits of the entries of $A$ are used for marking.
There is nothing wrong in such a bit-stealing technique, from a practical 
point\footnote{One of the authors of the current work also practiced it in 
a variant of the SamSAMi full-text index~\cite[Section~2.3]{GR17}.}, 
but those top bits may not always be available and thus in theory the space 
should be expressed as $O(q)$ words plus $O(n)$ bits.

We come back to the \textsf{ST-RMQ\textsubscript{CON}} algorithm.
As the name suggests, it builds the Sparse Table structure for the contracted array. 
All the queries can be answered in $O(q)$ time.
Interestingly, 
the construction and the queries are performed together, 
with re-use of the array storing the minima.
The ST construction time is $O(q\log q)$, 
but due to this clever trick, the size of the helper array is 
not $O(q \log q)$, but only $O(q)$.

\subsection{Block-based Sparse Table for the offline RMQ}
\label{sec:offline_our}

\subsubsection{BbST with the input array contraction}
\label{sec:offline_contraction}

On a high level, our first algorithm for the offline RMQ consists of 
the following four steps:

\begin{enumerate}
\item Sort the queries and remap them with respect 
to the contracted array's indices (to be obtained in step 2).
\item Contract $A$ to obtain $A_Q$ of size $O(q)$ (integers).
\item Build the block-based sparse table on $A_Q$ (see Section~\ref{sec:our1}), 
with blocks of size $k$.
\item Answer the queries, again in the manner of the solution 
from Section~\ref{sec:our1}.
\end{enumerate}

In the following paragraphs we are going to describe those steps in more detail, 
also pointing out the differences between our solution and Alzamel et al.'s one.

{\em (1) Sorting/remapping queries.}
Each of the $2q$ query endpoints is represented as a pair of 32-bit integers: 
its value (position in $A$) and its index in the query list $Q$.
The former 4-byte part is the key for the sort while the latter 4 bytes 
are satellite data.
In the serial implementation, we use kxsort\footnote{\url{https://github.com/voutcn/kxsort}}, 
an efficient MSD radix sort variant. 
In the parallel implementation, our choice was 
Multiway-Mergesort Exact variant implemented in GNU libstdc++ 
parallel mode library\footnote{\url{https://gcc.gnu.org/onlinedocs/libstdc++/manual/parallel_mode.html}}.
As a result, we obtain a sorted endpoint list $E[1 \dots 2q]$, 
where $E_i = (E_i^x, E_i^y)$ and $E_{i+1}^x \geq E_i^x$.
Alzamel et al. do not sort the queries, which is however possible 
due to marking bits in $A$.

{\em (2) Creating $A_Q$.}
Our contracted array $A_Q$ contains the minima of all areas 
$A[E_i^x \ldots E_{i+1}^x]$, in order of growing $i$.
$A_Q$ in our implementation contains thus (up to) $2q-1$ entries, twice less 
than in Alzamel et al.'s solution.
Like in the preceding solution, we also keep a helper array 
mapping from the indices of $A_Q$ into the original positions in $A$.

{\em (3) Sparse Table on blocks.}
Here we basically follow Alzamel et al. in their \textsf{ST-RMQ\textsubscript{CON}} variant, 
with the only difference that we work on blocks rather than individual elements of $A_Q$.
For this reason, this step takes 
$O(q + (q/k)\log(q/k)) = O(q(1 + \log(q/k) / k))$ time and $O((q/k)\log(q/k))$ 
space.
The default value of $k$, used in the experiments, is 512.

{\em (4) Answering queries.}
The {\em speculative reads} in the block-based sparse table 
(cf.~Section~\ref{sec:our1}) allow to answer a query often in constant time 
(yet, in rare cases an $O(k)$-time scan is needed).
This simple idea is crucial for the overall performance of our scheme.
In the worst case, we spend $O(k)$ time per query here, 
but on average, assuming uniformly random queries over $A$, 
the time is $O((k/q) \times k + (1-k/q) \times 1) = O(1 + k^2/q)$, 
which is $O(1)$ for $k = O(\sqrt{q})$.

Let us sum up the time (for a serial implementation) and space costs.
A scan over array $A$ is performed once, in $O(n)$ time.
The radix sort applied to our data of $2q$ integers from $\{1, \ldots, n\}$
takes (in theory) $O(q \max(\log n/\log q, 1))$ time.
Alternatively, introsort from C++ standard library (i.e., 
the std::sort function) would yield $O(q\log q)$ time.
To simplify notation, the $Sort(q)$ term will further be used 
to denote the time to sort the queries 
and we also introduce $q' = q/k$.
$A_Q$ is created in $O(q)$ time.
Building the Sparse Table on blocks adds $O(q + q'\log q')$ time.
Finally, answering queries requires $O(qk)$ time in the worst case 
and $O(q + k^2)$ time on average.
In total, we have $O(n + Sort(q) + q'\log q' + qk)$ time in the worst case.
The extra space is $O(q'\log q')$.

In Section~\ref{sec:our1}
we presented a variant of the Sparse Table with arity higher than two
for the online RMQ problem.
Now we discuss it in the context of the offline RMQ.
The worst-case time of handling $q$ queries becomes 
$O(n + Sort(q) + q'\log q'/\log\ell + q\ell + qk)$, 
which is minimized for 
$\ell = \max(\log q' / (k\log\log q'), 2)$.
With $k$ small enough to have $\ell = \log q' / (k\log\log q')$, 
we obtain
$O(n + Sort(q) + q'\log q' / \log\log q' + qk)$ overall time 
and the required extra space is $O(q'\log q' / \log\log q')$ words.

If we focus on the average case, where the last additive term 
of the worst-case time turns into $k^2/q$, 
it is best to take $k = \sqrt{q}$, which implies $\ell = 2$.
In other words, this idea has its niche only considering 
the worst-case time, 
where for a small enough $k$ both the time and the space 
of the standard block-based Sparse Table solution are improved.

\subsubsection{BbST with no input array contraction}
\label{sec:offline_no}

The simple solution presented in Section~\ref{sec:our1}, due to a very fast 
construction, seems to be suitable also for the offline RMQ problem.
This variant greatly simplifies the procedure described in 
Section~\ref{sec:offline_contraction}, as now there is no need to 
sort the queries.
Basically, we reduce the previous variant to the last two stages.
Naturally, this comes at a price: the extra space usage becomes 
$O((n/k)\log(n/k))$ words 
(yet the optimal choice of $k$ may be different, closer to $\sqrt{n}$), 
but query times often become very competitive.
%

Let us focus on the space and time complexities for this variant, 
for both the worst and the average case.
The analysis resembles the one for the variant with the contracting of $A$.
We have two parameters, $n$ and $k$, and two stages of the algorithm.
The former stage takes $O(n + (n/k)\log(n/k))$ time, 
the latter takes $O(qk)$ time in the worst case and 
$O(q(1 + k^2/n))$ on average (which is $O(q)$ if $k = O(\sqrt{n})$).
In total we have $O(n + (n/k)\log(n/k) + qk)$ time in the worst case
and $O(n + (n/k)\log(n/k) + q)$ time on average, 
provided in the latter case that $k = O(\sqrt{n})$.
The space, expressed in words, 
is $O((n/k)\log(n/k))$.
To minimize both the time and the space for the average case 
we set $k = \Theta(\sqrt{n})$.
Then the average time becomes $O(n + \sqrt{n}\log\sqrt{n} + q) = O(n + q)$ 
and the space is $O(\sqrt{n}\log n)$.

\section{Experimental results}

\begin{figure}[pt]
\centerline{
\includegraphics[width=0.495\textwidth,scale=1.0]{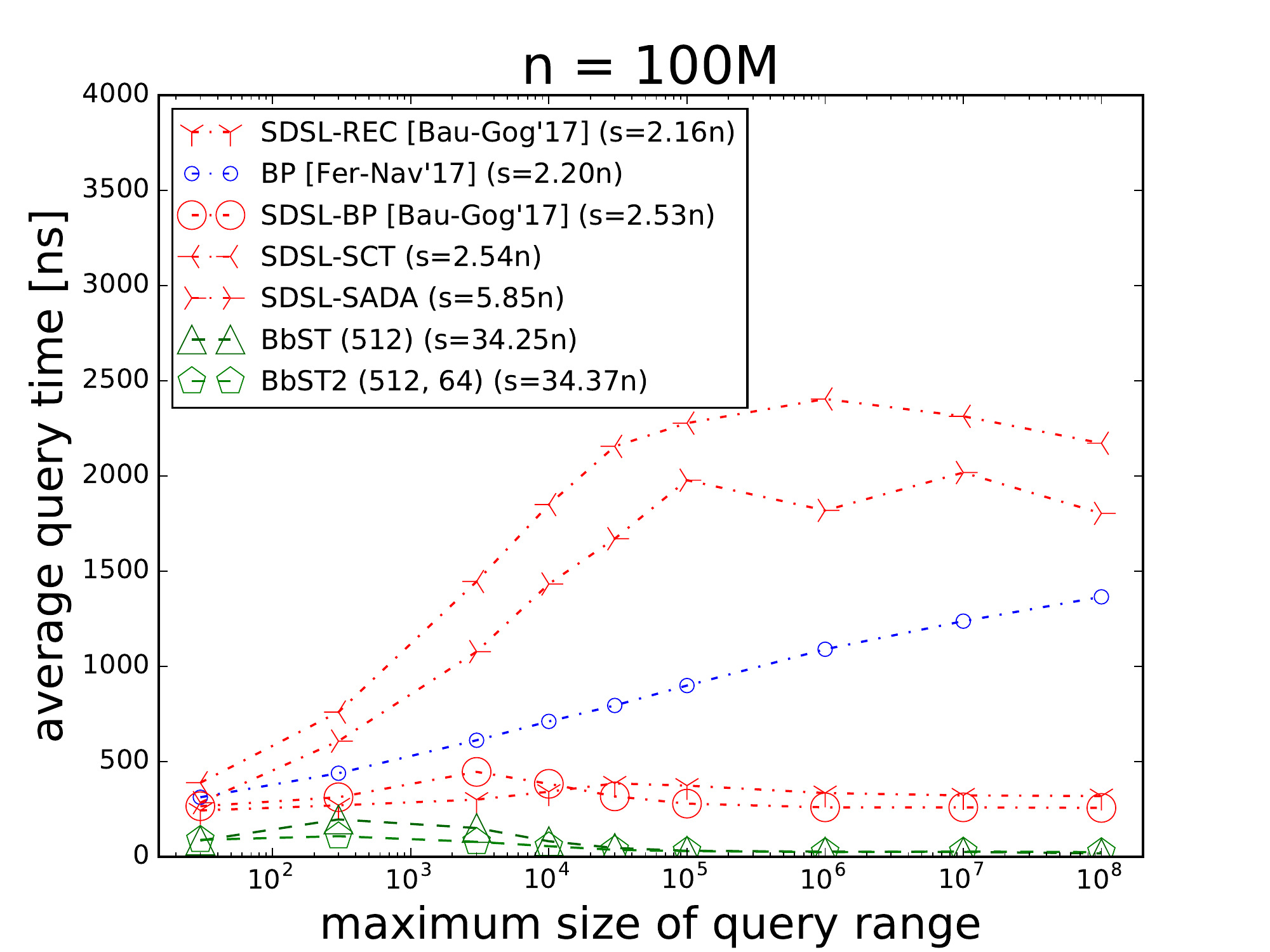}
\includegraphics[width=0.495\textwidth,scale=1.0]{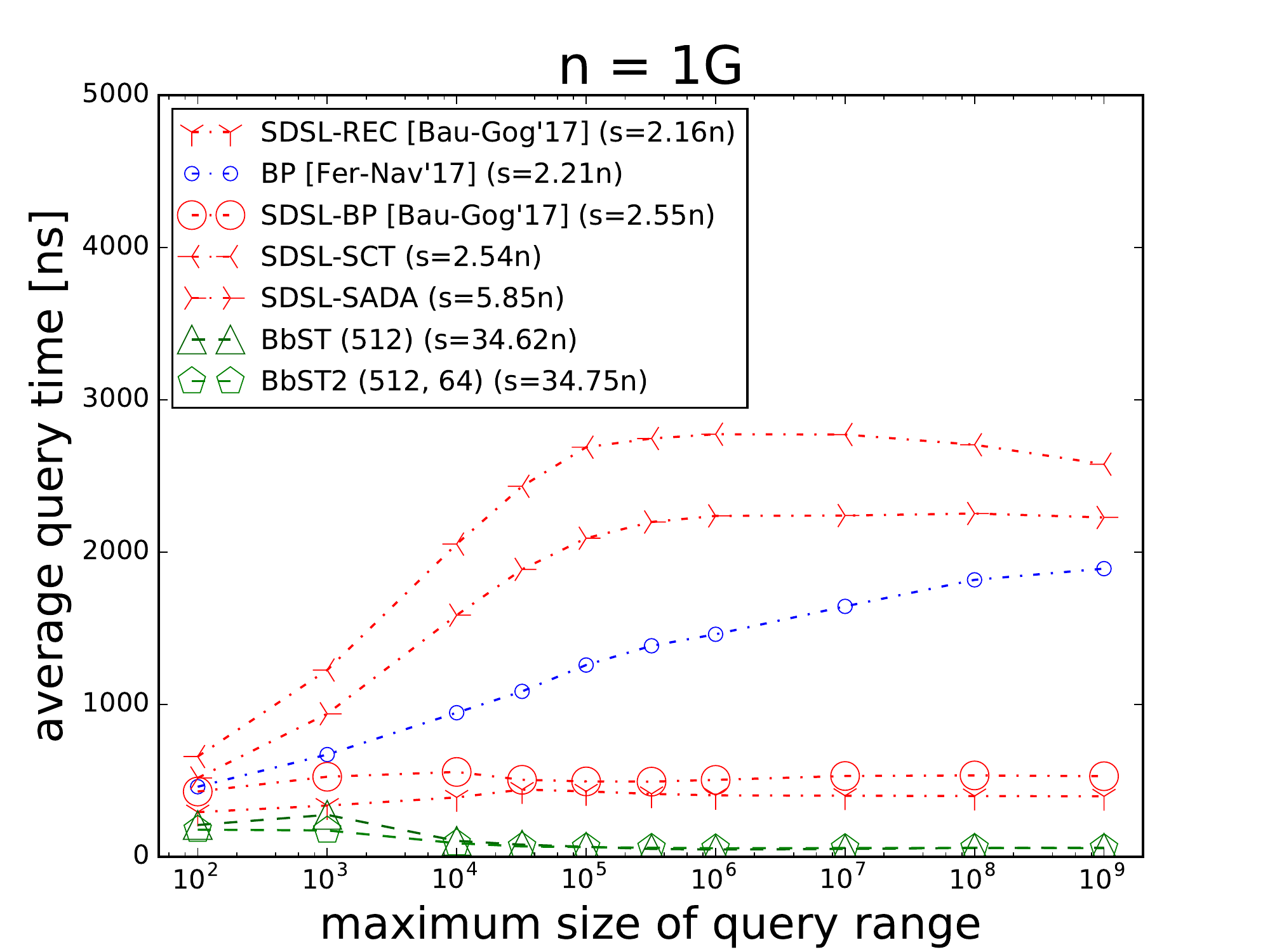}
}
\centerline{
\includegraphics[width=0.495\textwidth,scale=1.0]{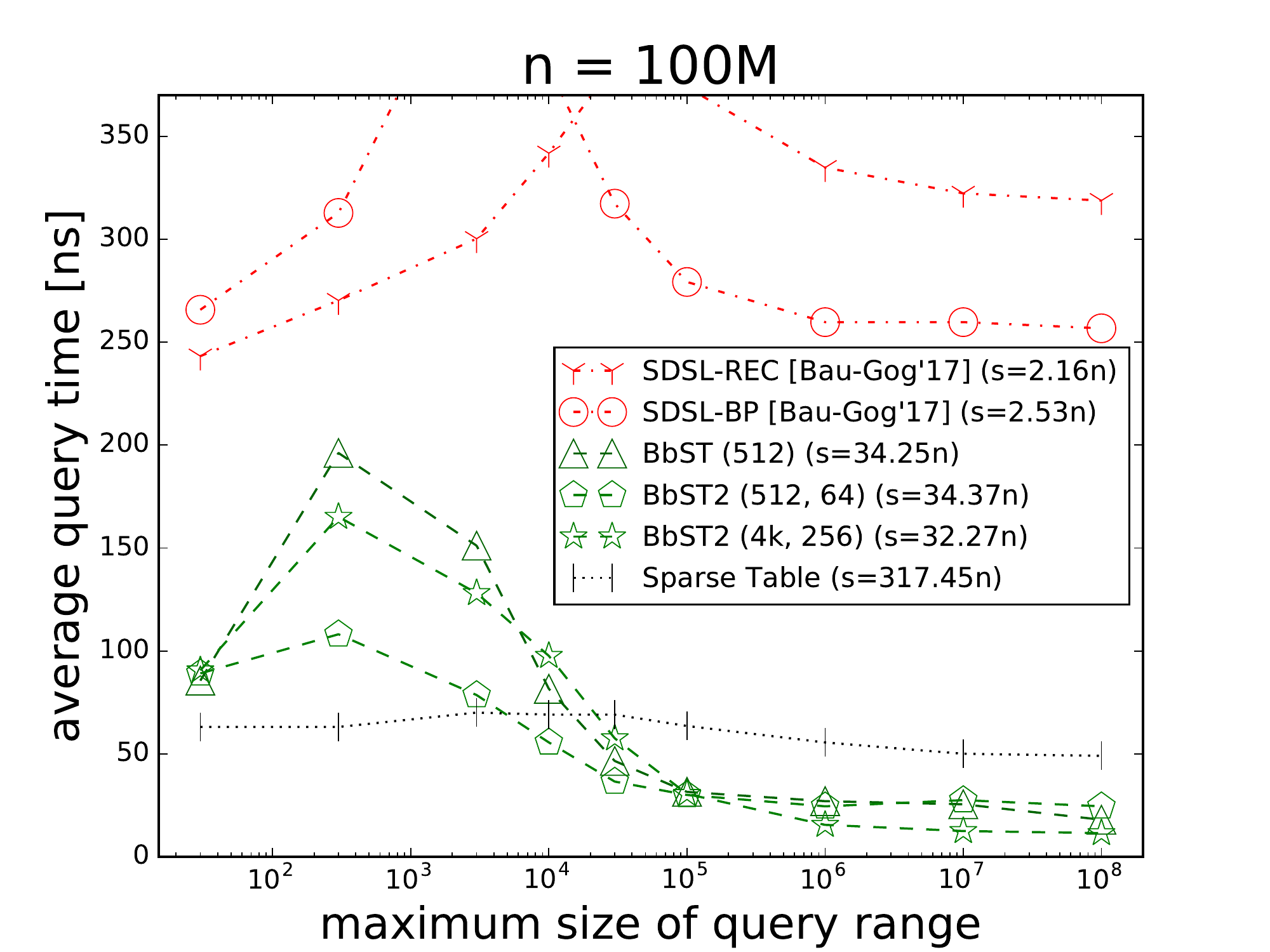}
\includegraphics[width=0.495\textwidth,scale=1.0]{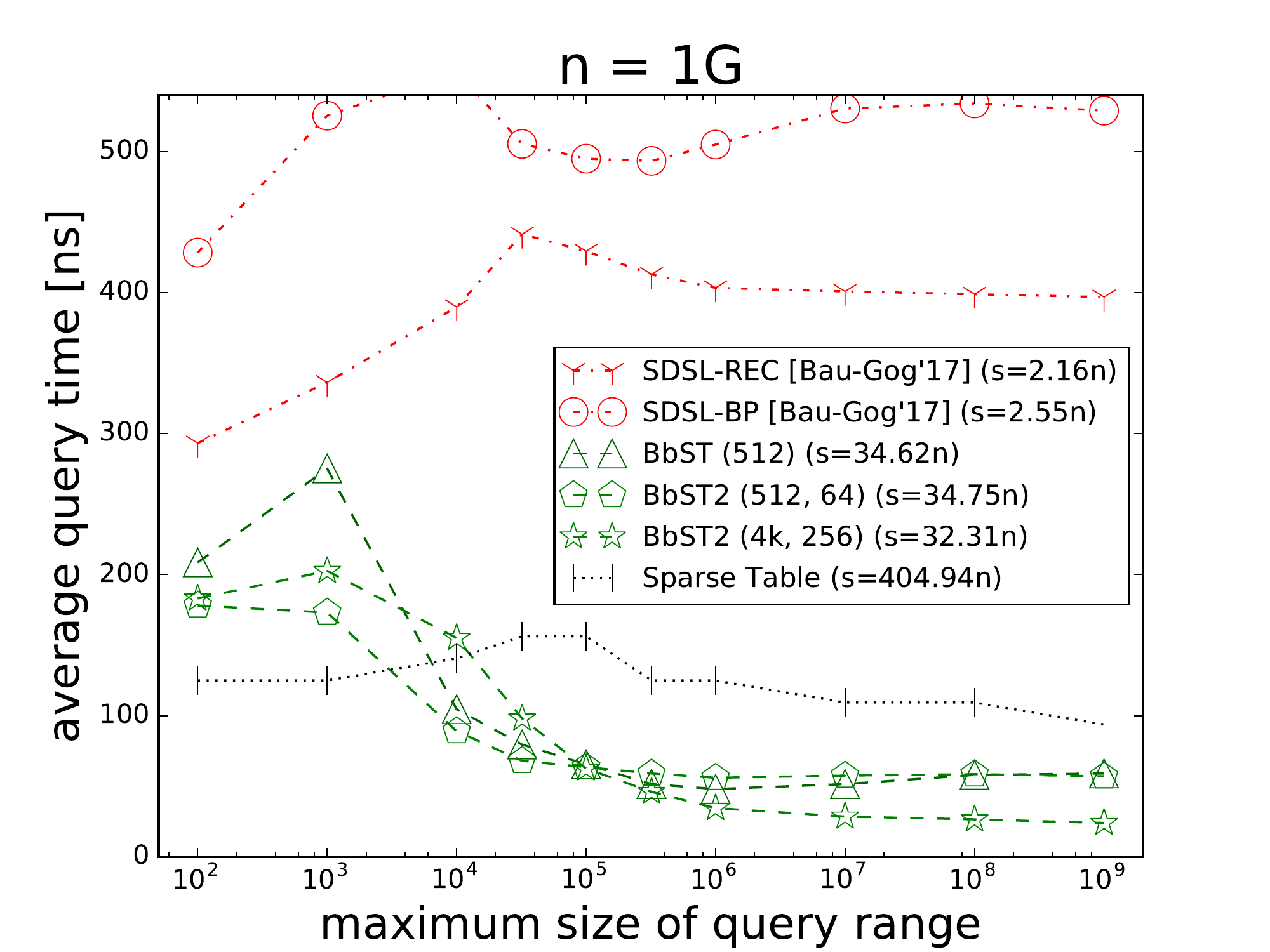}
}
\caption[Fig1a]
{Average query times for ranges of varying maximum width 
(uniformly random from 1 to the given value)
and two sizes of the input sequence (100M and 1G). 
For each maximum width 1M queries were used.
The space usage, in bits, for particular solutions is given in the legends,  
in parentheses.
The bottom figures include only the faster solutions from the top figures, 
plus a line for the standard Sparse Table.}
\label{fig:1a}
\end{figure}

\begin{figure}[pt]
\centerline{
\includegraphics[width=0.495\textwidth,scale=1.0]{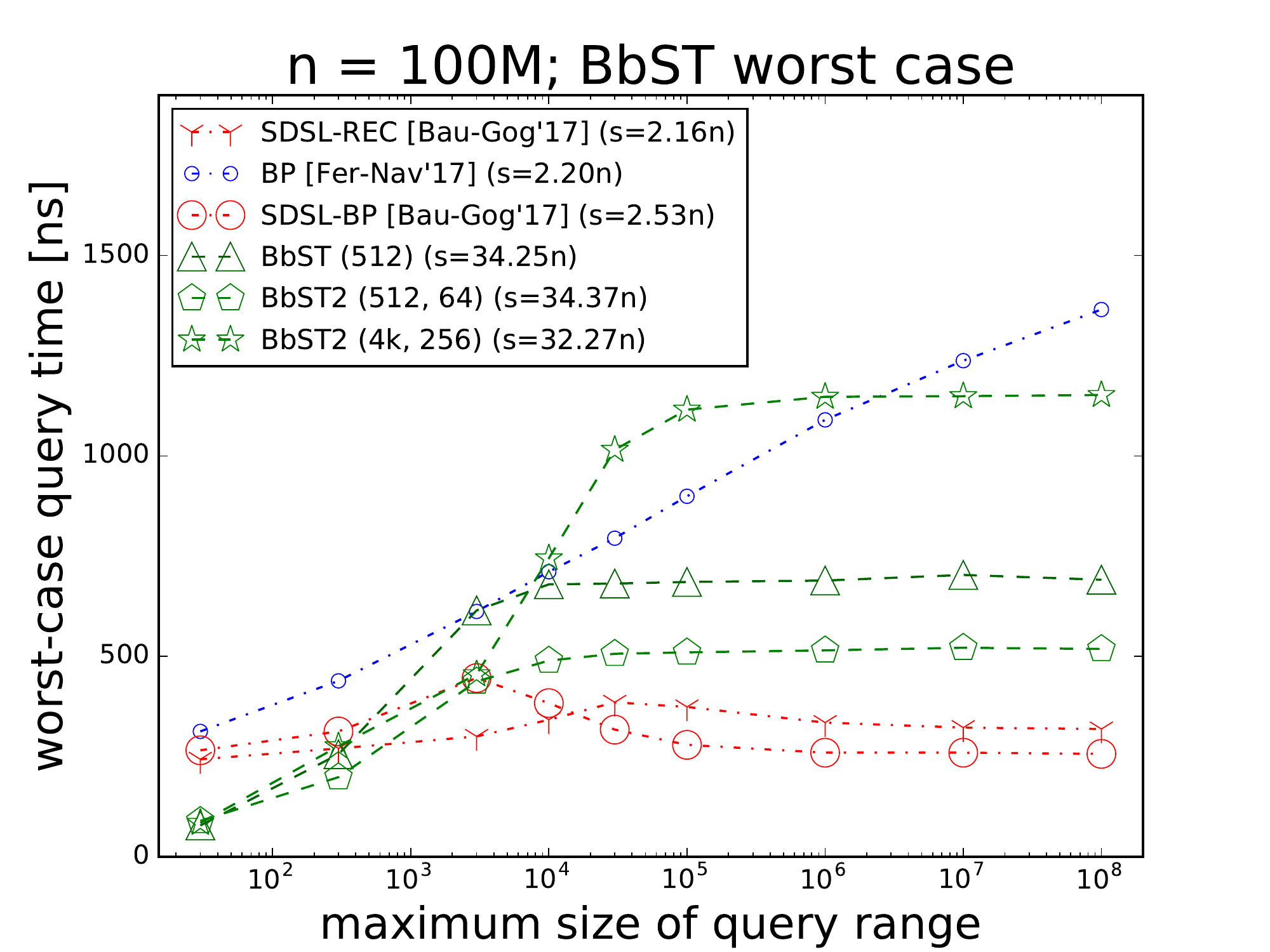}
\includegraphics[width=0.495\textwidth,scale=1.0]{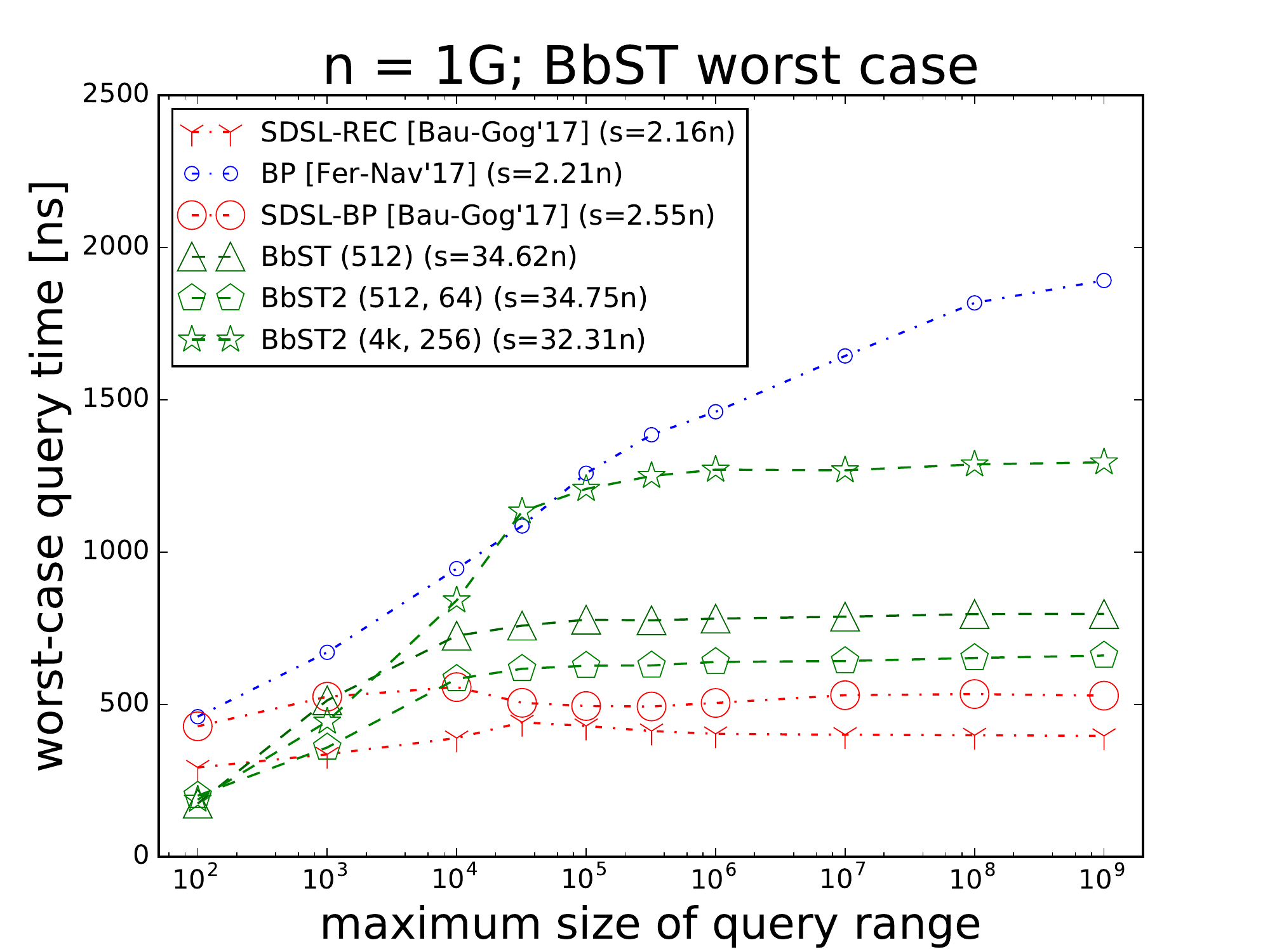}
}
\caption[Fig1b]
{Estimated worst-case query times for ranges of varying maximum width 
(uniformly random from 1 to the given value) and two sizes of the input sequence 
(100M and 1G). 
For each maximum width 1M queries were used.
The space usage, in bits, for particular solutions is given in the legends, 
in parentheses.}
\label{fig:1b}
\end{figure}

For the experiments, we use the array $A$ storing 
random 32-bit unsigned integers.
The queries are pairs of the form $(\ell_i, r_i)$, 
where $\ell_i$ is drawn uniformly random from the whole sequence 
and $r_i - \ell_i$ is between 0 and a specified range width limit.

Our algorithms were implemented in C++ and compiled with 32-bit gcc 7.2.0 
with \texttt{-O3 -mavx2} switches.
The source codes can be downloaded from \url{https://github.com/kowallus/BbST}.
The experiments were conducted on a desktop PC equipped with 
a 4-core Intel i7 4790 3.6\,GHz CPU and 32\,GB of 1600\,MHz DDR3 RAM (9-9-9-24), 
running Windows 10 Professional.
All presented timings in all tests are medians of 7 runs, 
with cache flushes in between.

\begin{figure}[pt]
\centerline{
\includegraphics[width=0.495\textwidth,scale=1.0]{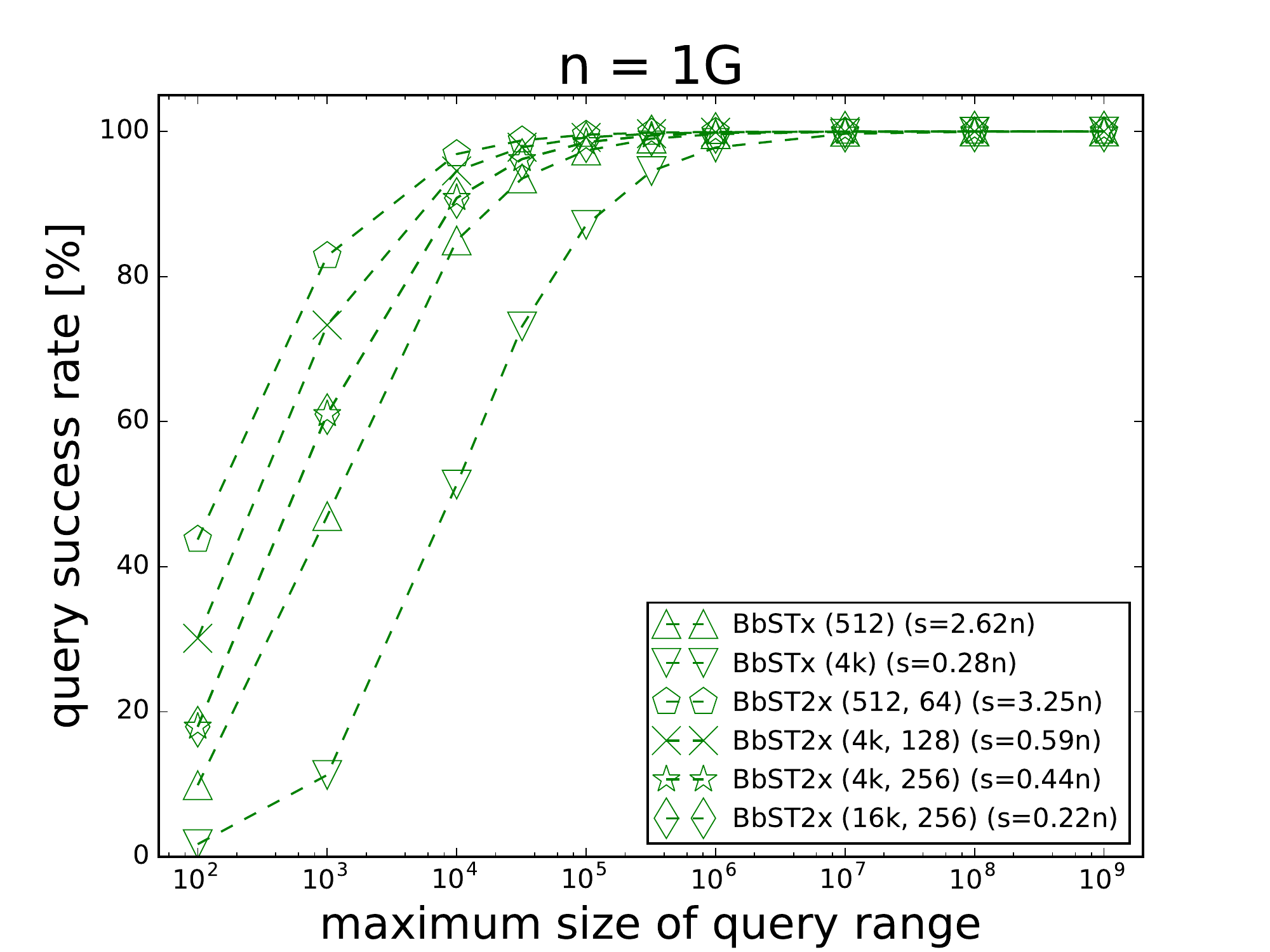}
\includegraphics[width=0.495\textwidth,scale=1.0]{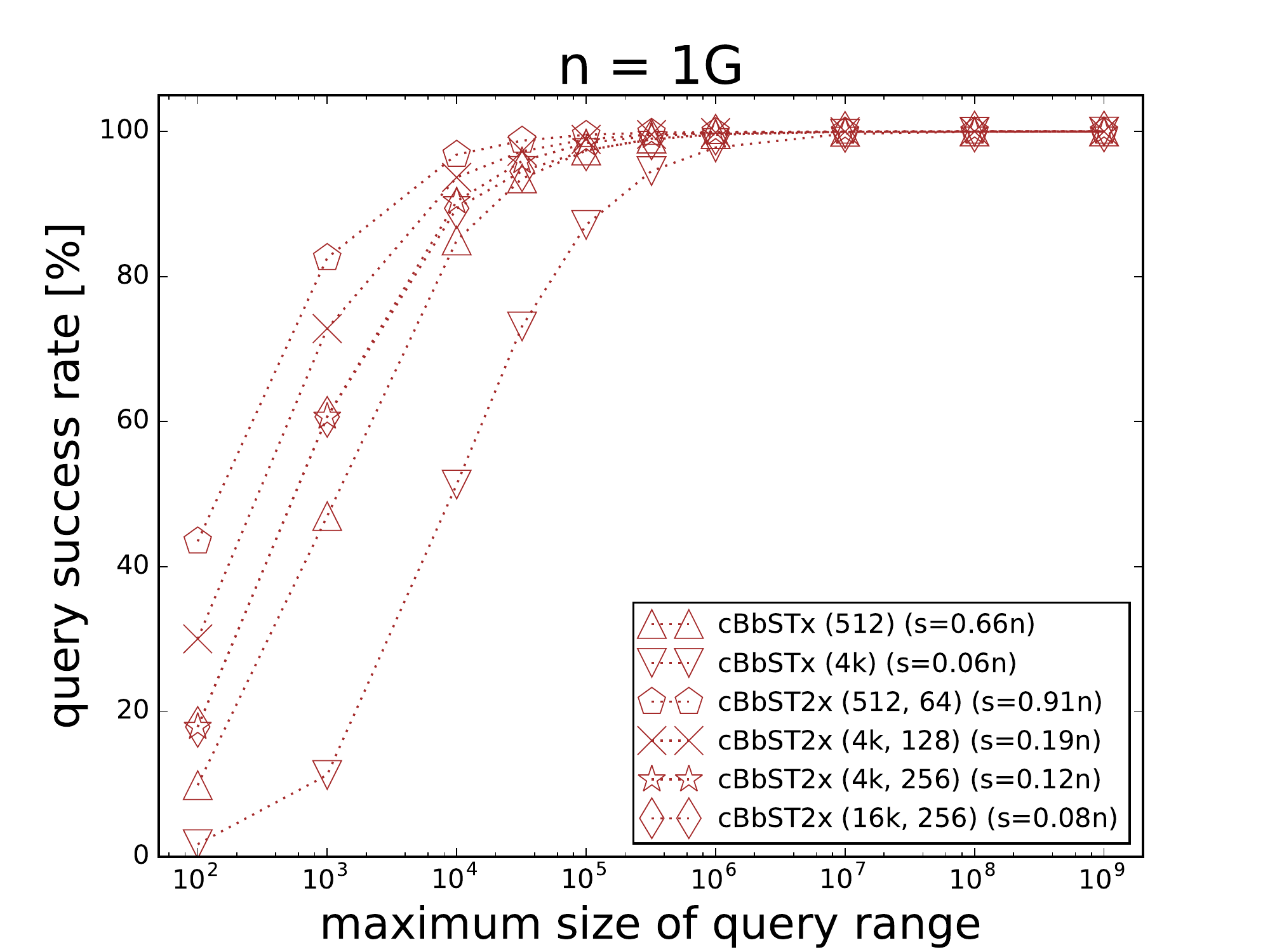}
}
\caption[Fig2]
{The query success rate, i.e., how often a random query can be handled by our data 
structure without accessing the input array $A$.
For each maximum width 1M queries were used.
The space usage, in bits, for particular solutions is given in the
legends, in parentheses.
Note that these solutions are not full RMQ-answering algorithms.}
\label{fig:2}
\end{figure}

We start with the experiments in the traditional, offline, RMQ scenario.
Fig.~\ref{fig:1a} presents average query times in function of growing maximum 
query range size.
We use the following algorithms in the comparison:

\begin{itemize}
\item \textsf{SDSL-SCT} and \textsf{SDSL-SADA}, two RMQ implementations from the well-known 
SDSL library~\cite{GogBMP14} (\url{https://github.com/simongog/sdsl-lite}),

\item \textsf{BP} (Balanced Parentheses) algorithm by Ferrada and Navarro~\cite{FN16} 
(\url{https://github.com/hferrada/rmq.git}),

\item \textsf{SDSL-BP} and \textsf{SDSL-REC}, 
two algorithms by Baumstark et al.~\cite{BaumstarkGHL17} 
(\url{https://github.com/kittobi1992/rmq-experiments}),

\item \textsf{BbST}, our baseline solution, with block size of $k = 512$,

\item \textsf{BbST2}, our two-level solution, with block sizes $(k_1, k_2)$ 
set to $(512, 64)$ or $(4096, 256)$, respectively.
\end{itemize}

\begin{figure}[pt]
\centerline{
\includegraphics[width=0.495\textwidth,scale=1.0]{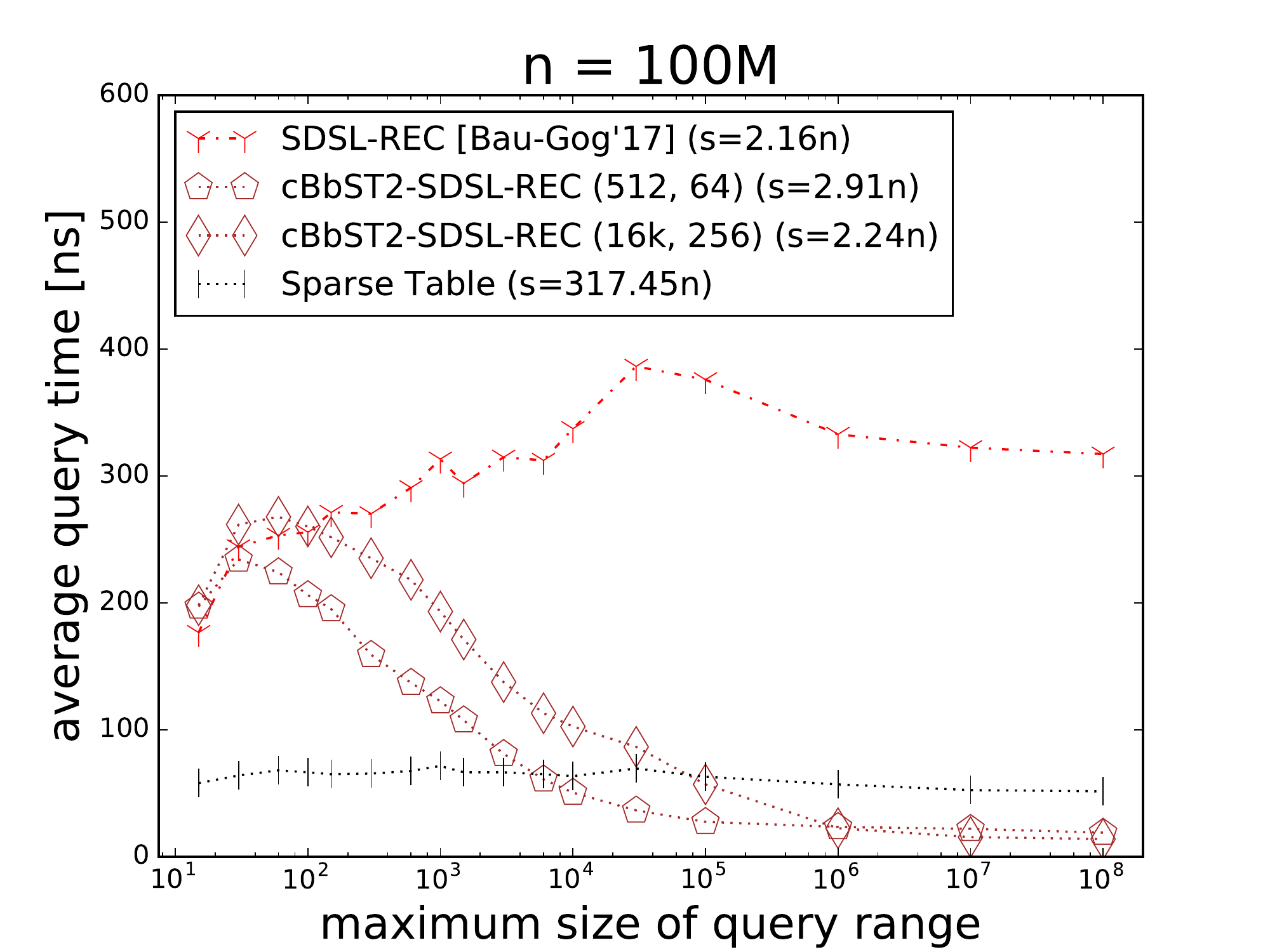}
\includegraphics[width=0.495\textwidth,scale=1.0]{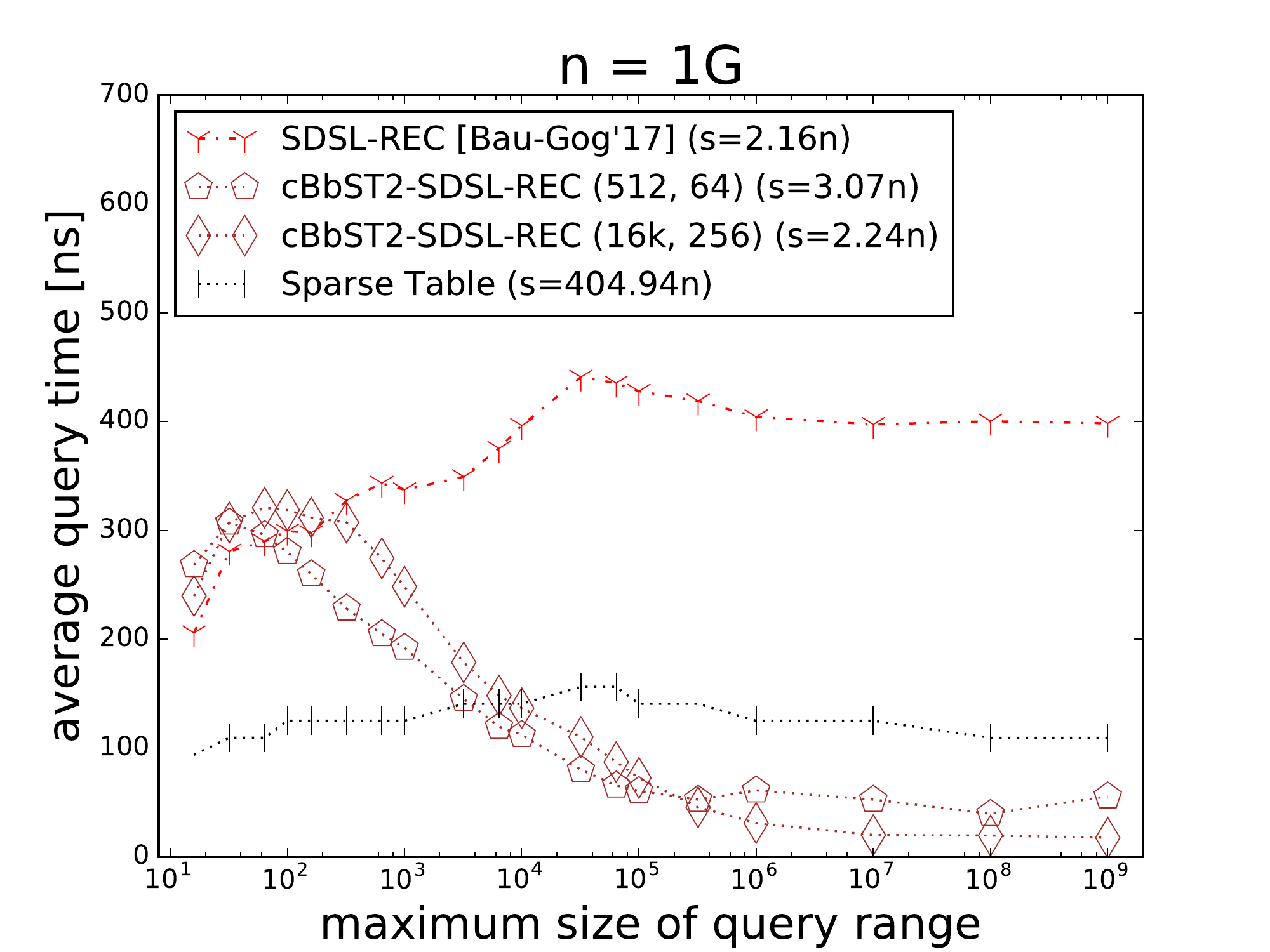}
}
\caption[Fig3]
{Average query times for ranges of varying maximum width 
(uniformly random from 1 to the given value)
and two sizes of the input sequence (100M and 1G). 
Our hybrid, \textsf{cBbST2-SDSL-REC}, in two parameter configurations, 
is compared against the fastest non-hybrid solution from the literature, 
\textsf{SDSL-REC}.
For each maximum width 1M queries were used.
The space usage, in bits, for particular solutions is given in the legends,  
in parentheses.}
\label{fig:3}
\end{figure}

The numbers is parentheses give the space usage in bits. 
Note that our algorithms require access to array $A$, which results in the overhead of 
$32 n$ bits.
The input size $n$ is 100 million in the left figures and 1 billion in the right ones.
The top figures present all lines while in the bottom ones we focus on the faster ones 
(note the time scale).
The line with vertical markers
stands for the classical Sparse Table solution; it shows the performance 
of this very simple data structure, which is unbeatable for relatively narrow queries 
(of width up to a few thousands), yet the required space is around 320--400$n$ bits.

We can see that our idea of using blocks not only reduces the ST space by an order of 
magnitude, but also speeds up queries on wide intervals.
\textsf{SDSL-REC} is the most succinct solution.
\textsf{BP} is a close second in space, yet not so fast as \textsf{SDSL-BP} 
and \textsf{SDSL-REC}.
Of these two, \textsf{SDSL-REC} seems to be the method of choice, 
even if not always faster than \textsf{SDSL-BP}.
The performance of our variants usually improves with growing range widths, 
which is not the case of the competitors.
The two-level variant, \textsf{BbST2}, is more succinct and also usually faster 
than \textsf{BbST}.
We note that \textsf{BbST2} is, roughly speaking, about twice faster than \textsf{SDSL-REC}
with range widths up to a few thousands (the gap is greater than 2-fold for 
$n = 100\textrm{M}$ and smaller for $n = 1\textrm{G}$), 
yet grows to an order of magnitude for wide queries.

In Fig.~\ref{fig:1b} we {\em estimate} the worst-case, rather than average, query time
of our algorithms.
In this experiment, for each query we scan the two blocks to which its boundaries belong
(no matter if this scan were really needed) and the averages over such times
are presented.
Note that a `direct' measurement of the worst case, that is, taking the maximum time 
over many queries, is hard to perform reliably, as the times are below 1\,$\mu$s.
As expected, in this comparison our algorithms are not really competitive, 
except for narrow ranges (maximum width of 30 in the test).
Yet, for much wider ranges BbST variants are inferior in speed only to 
\textsf{SDSL-REC} and \textsf{SDSL-BP}.
Interestingly, the query times of \textsf{BP} grow roughly linearly in the logarithm 
of the range width; for the other tested algorithms the timings stabilize.

For the experiments to follow in most cases we present the results only 
for $n = 1\textrm{G}$, to save space (in the case of $n = 100\textrm{M}$ 
the trends are similar).

Our next attempt was to combine the block-based sparse table with \textsf{SDSL-REC}, 
in order to get rid of the input array during the query handling.
The variants with letter `x' in their names, shown in Fig.~\ref{fig:2}, 
are {\em not} yet hybrids; they do not answer RMQ in all cases. 
They simply get rid of array $A$ and are thus unable to scan over an interval.
If the precomputed minima are not enough to answer a given query, 
the algorithm \textsf{(c)BbSTx} (resp. \textsf{(c)BbST2x}) is unable to given an answer.
The query success rate tells how often the query can be handled. 
Note that now the space is much reduced.
The left figure presents variants based on the standard \textsf{BbST(2)}, 
while the right one shows their compact versions, with prefix `c' in their names.
As expected, the compact variants require less space, but their query success rates 
overlap with the values for the corresponding non-compact variants.

For the hybrids involving \textsf{cBbST2}, 
we used the following formula for quantizing the block minimum values 
for the blocks of size $k_2$: 
\begin{equation*}
\lfloor maxQ \times (1 - (maxMin - v)^{8} / (maxMin - minMin)^{8}) \rfloor,
\end{equation*}
where $v$ is a block minimum value, 
and $maxMin$ (resp. $minMin$) is the largest (resp. smallest) minimum 
among the minima for blocks of size $k_2$.
The formula was found experimentally.

As the compact variants are recommended both for speed and space frugality, 
we combined them with 
\textsf{SDSL-REC}
variants into hybrids (Fig.~\ref{fig:3}).
We can see that for wide intervals our hybrids are faster than 
\textsf{SDSL-REC}
by 
more than an order of magnitude, while for narrow ones (up to a few hundred 
in width) the gap is quite narrow.
Yet, the more successful of our variants, the hybrid with block sizes of 
16384
and 256, respectively, is defeated in speed (by about 10\%) only for 
the narrowest interval.
Fortunately, the same variant is more compact of our two, 
with 2.24$n$ bits of space, which is not much more than 2.16$n$ bits 
of 
\textsf{SDSL-REC}.


An important facet of every data structure is its construction time.
Table~\ref{table:build} presents the construction times (and space usage) 
for several RMQ algorithms or their configurations, 
for the input array of size $n = 1\textrm{G}$.
We can see that the plain \textsf{BbST} is clearly the fastest, 
about 40 times faster than the fastest solution with constant worst case time queries, 
\textsf{SDSL-SCT}.
Note also that in the construction time for \textsf{SDSL-SCT} over
$1\textrm{G}$ elements we can build \textsf{BbST} and answer
from about $100\textrm{M}$ to $400\textrm{M}$ queries.
Our two-level variant, \textsf{BbST2}, is still very fast in construction.
The hybrids, however, must require more time to build than \textsf{SDSL-REC}, 
which is their component.
\textsf{ST}, as clearly the most memory-demanding data structure, 
is also the slowest to build.

\begin{table}[t!]
\centering
\begin{tabular}{lrr}
\hline
variant~~~~~ & build time~~~~& size / $n$ \\
             & [s]       ~~~~& [bits]~\\
\hline
\textsf{SDSL-SADA}                    &  212.5~~~~&   5.85 \\
\textsf{SDSL-SCT}                     &   23.9~~~~&   2.54 \\
\textsf{BP}                           &   66.6~~~~&   2.21 \\
\textsf{SDSL-BP}                      &   26.0~~~~&   2.55 \\
\textsf{SDSL-REC}                     &   62.6~~~~&   2.16 \\
\textsf{ST}                           &  436.6~~~~& 404.94 \\
\textsf{BbST}, $k = 512$              &    0.6~~~~&  34.63 \\
\textsf{BbST2}, $k_1 = 512$, $k_2 = 64$ &  2.7~~~~&  34.75 \\
\textsf{BbST2}, $k_1 = 4096$, $k_2 = 256$ & 2.8~~~~& 32.31 \\
\textsf{cBbST-SDSL-REC}, $k = 512$    &   66.0~~~~&   2.82 \\
\textsf{cBbST2-SDSL-REC} $(512, 64)$  &   67.6~~~~&   3.07 \\
\textsf{cBbST2-SDSL-REC} $(16384, 256)$ & 67.6~~~~&   2.24 \\
\hline
\end{tabular}
\vspace{4mm}
\caption{Construction times and space usage for several RMQ algorithms, 
for $n = 1\textrm{G}$.
All implementations are single-threaded.}
\label{table:build}
\end{table}

Table~\ref{table:components} focuses on the space usage for individual components 
of our variants.
The values in column ``backend RMQ data'' are equal to either 32 
(i.e., the size of input data in array $A$) 
for non-hybrid solutions 
or the number of bits per element spent in \textsf{SDSL-REC}.
The size of the sparse table component is relatively large for $k_1 = 512$, 
but improves in the compact variants (those with `c' in their names) 
and, of course, gets reduced with growing $k_1$.
The overhead of the second level blocks is given in the last column.

\begin{table}[t!]
\centering
\begin{tabular}{lrrr}
\hline
variant~~~~~ & backend~~~~& sparse~~~~& second \\
             & RMQ data~~~~& table~~~~& level \\
\hline
\textsf{BbST}, $k = 512$              &     32~~~~&  2.63~~~~& --- \\
\textsf{BbST2}, $k_1 = 512$, $k_2 = 64$ &   32~~~~&  2.63~~~~& 0.13 \\
\textsf{BbST2}, $k_1 = 4096$, $k_2 = 256$ & 32~~~~&  0.28~~~~& 0.03 \\
\textsf{BbST-SDSL-REC}, $k = 512$     &   2.16~~~~&  2.63~~~~& --- \\
\textsf{BbST2-SDSL-REC} $(512, 64)$  &    2.16~~~~&  2.63~~~~& 0.63 \\
\textsf{BbST2-SDSL-REC} $(4096, 256)$  &  2.16~~~~&  0.28~~~~& 0.16 \\
\textsf{cBbST-SDSL-REC}, $k = 512$    &   2.16~~~~&  0.66~~~~& --- \\
\textsf{cBbST2-SDSL-REC} $(512, 64)$  &   2.16~~~~&  0.66~~~~& 0.25 \\
\textsf{cBbST2-SDSL-REC} $(16384, 256)$ & 2.16~~~~&  0.01~~~~& 0.06 \\
\hline
\end{tabular}
\vspace{4mm}
\caption{Space usage for individual data structure components. 
All numbers are in bits per element.}
\label{table:components}
\end{table}

The last experiments concerned the offline RMQ scenario.
Here, a batch of $q$ queries is handled, where $q \ll n$.
The comparison comprises the following algorithms:

\begin{itemize}
\item \textsf{ST-RMQ\textsubscript{CON}}, by Alzamel et al.~\cite{ACIP17}
(\url{https://github.com/solonas13/rmqo}),

\item \textsf{BbST\textsubscript{CON}}, a version of our block-based sparse table 
with contracted input array, with block size of $k = 512$,

\item \textsf{BbST} and \textsf{BbST2}, our algorithms used in the previous 
experiments, with block sizes set to $k = 512$ (\textsf{BbST}) 
and $(k_1, k_2) = (4096, 256)$ (\textsf{BbST2}).
\end{itemize}

\begin{figure}[pt]
\centerline{
\includegraphics[width=0.495\textwidth,scale=1.0]{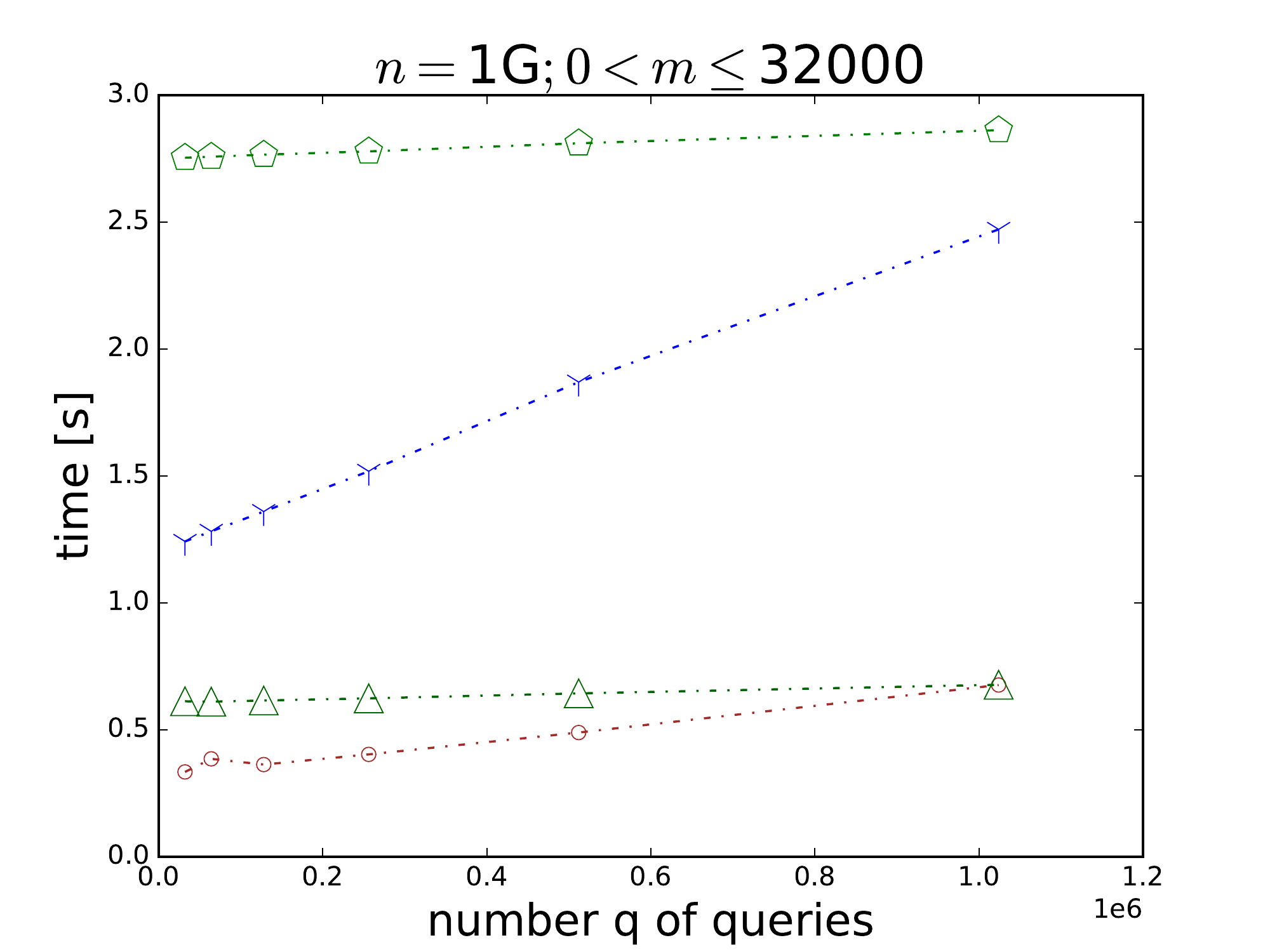}
\includegraphics[width=0.495\textwidth,scale=1.0]{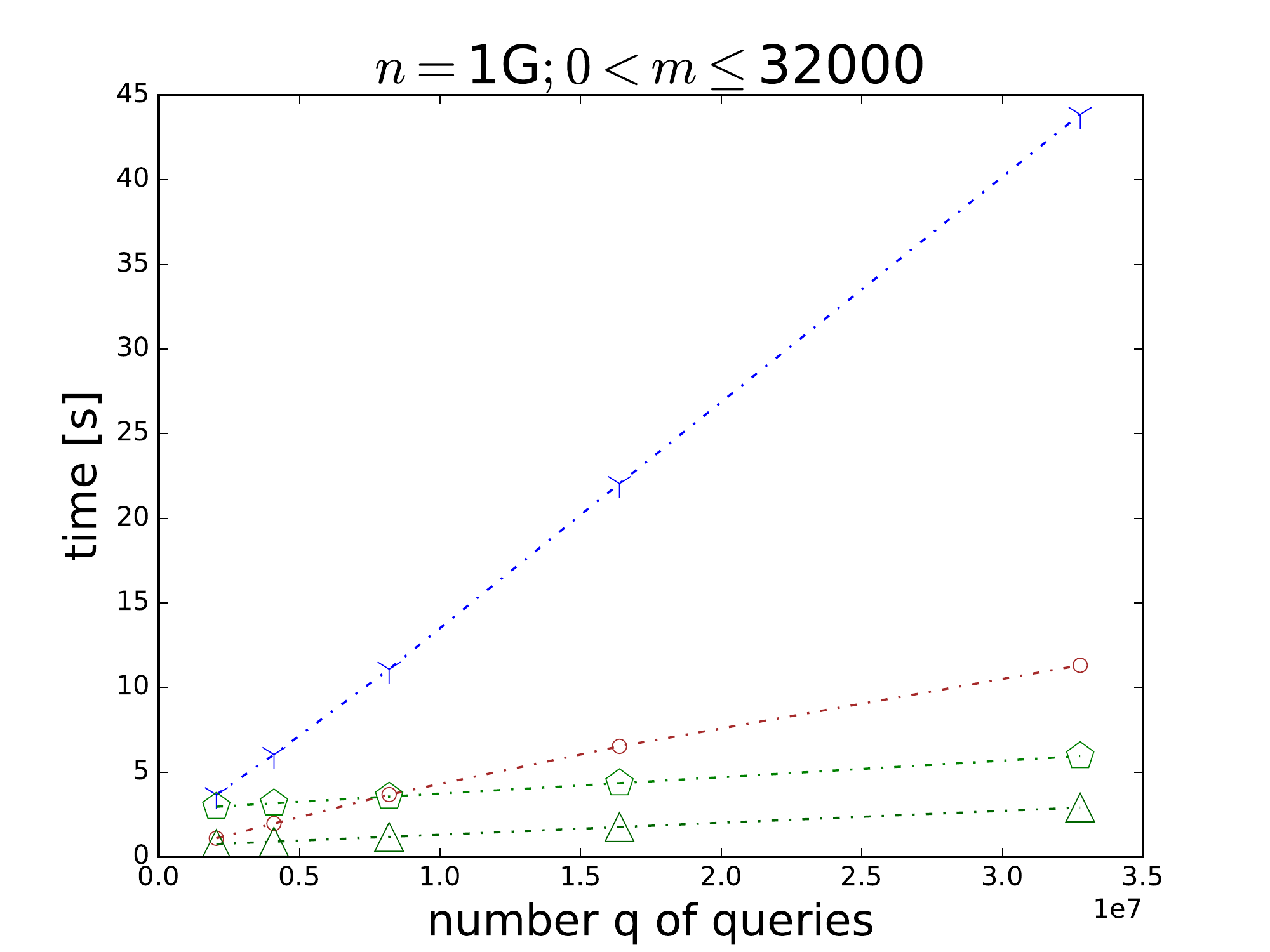}
}
\centerline{
\includegraphics[width=0.495\textwidth,scale=1.0]{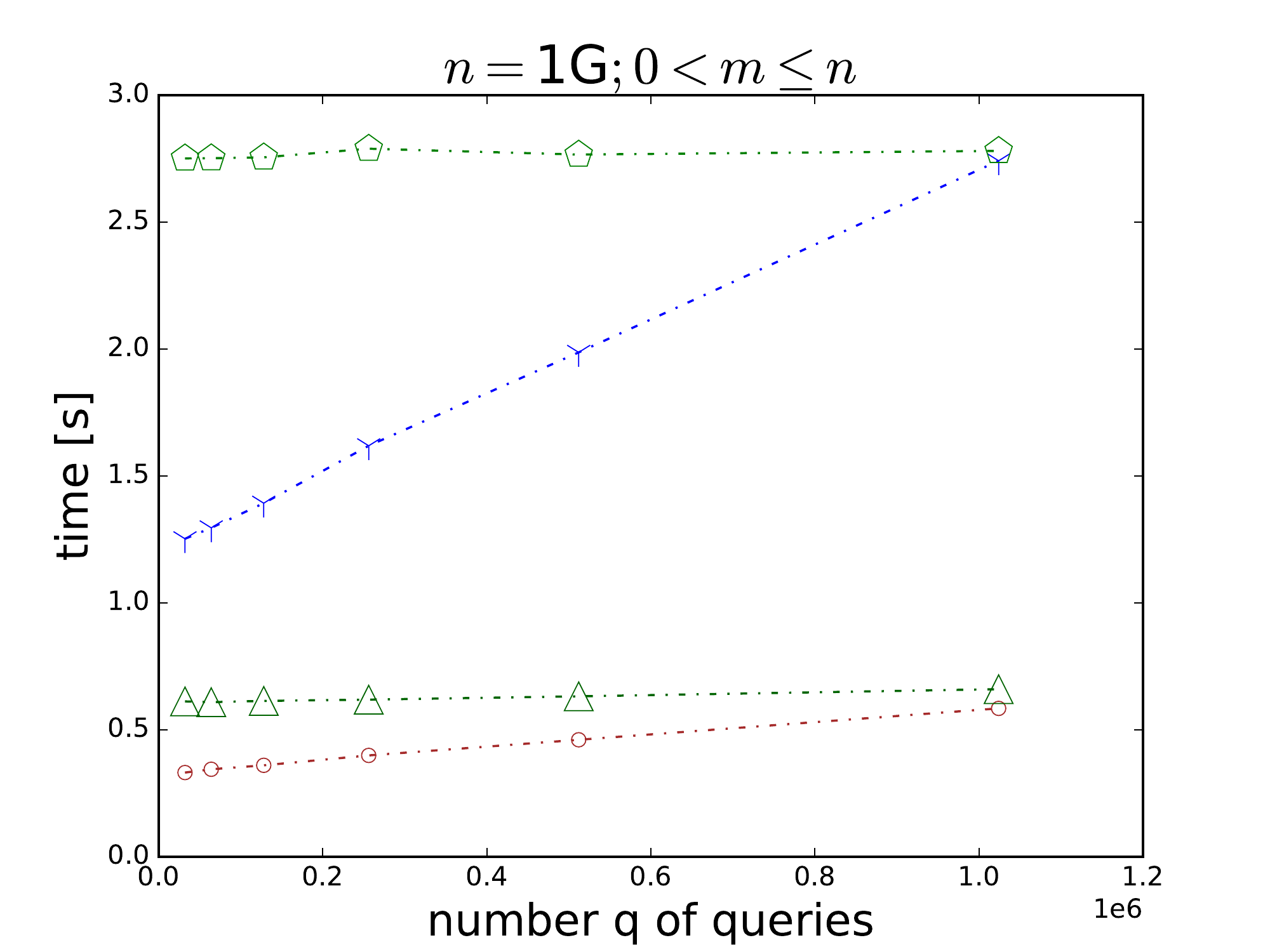}
\includegraphics[width=0.495\textwidth,scale=1.0]{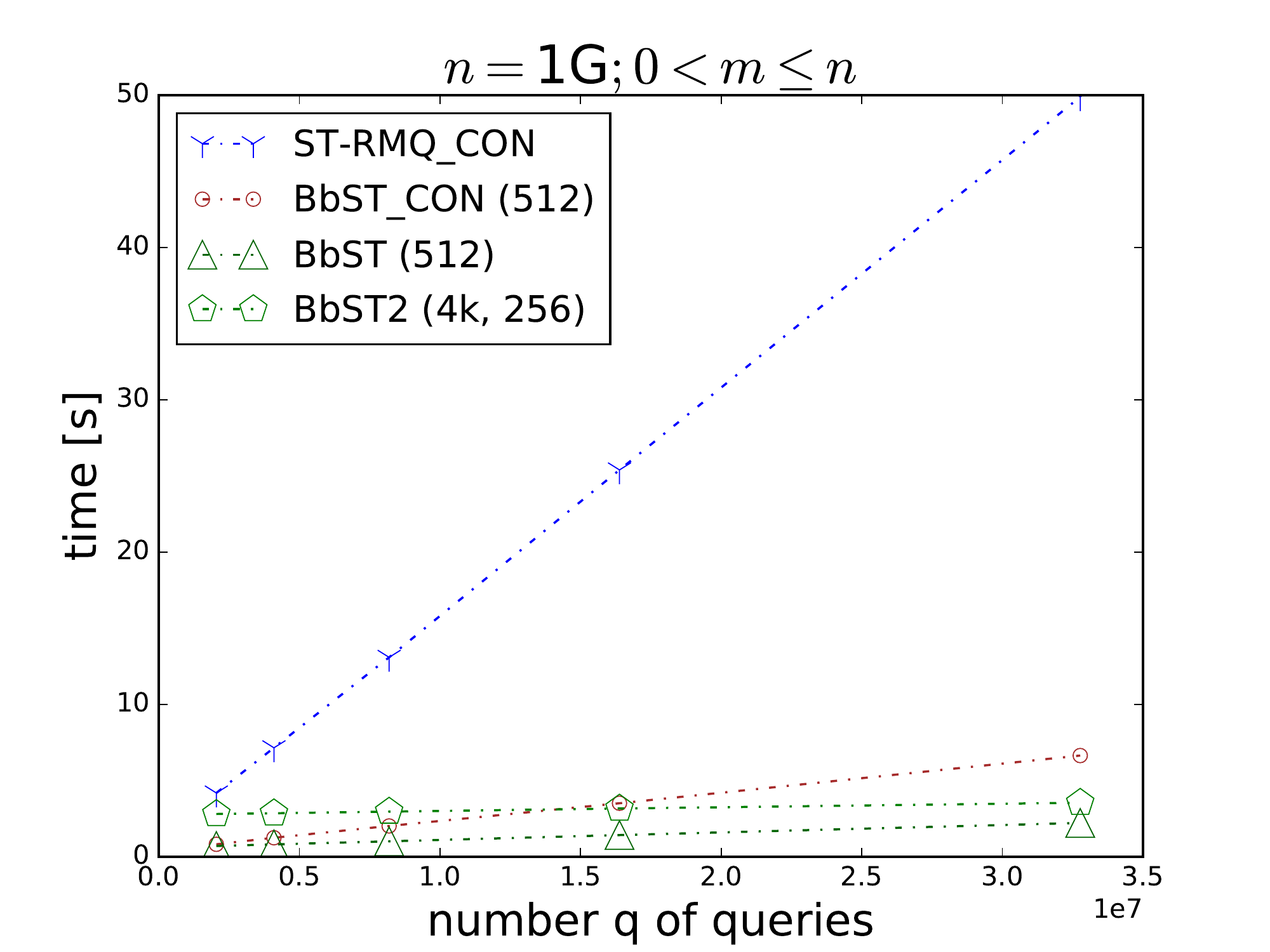}
}
\caption[Fig4]
{Running times for with varying number of queries $q$, 
from $\sqrt{n}$ to $32\sqrt{n}$ (left figures) 
and from $64\sqrt{n}$ to $1024\sqrt{n}$ (right figures), 
where $n = 1\textrm{G}$.
The symbol $m$ denotes a query width.
In the top figures the maximum query width is 32K, while in the bottom ones 
it is $1\textrm{G}$.}
\label{fig:4}
\end{figure}

We can see (Fig.~\ref{fig:4}) that the relative advantage of our variants
over \textsf{ST-RMQ\textsubscript{CON}} grows with the number of queries.
In any case, our algorithm is several times faster than its predecessor.
For small enough $q$ (the left figures), 
\textsf{BbST\textsubscript{CON}} dominates over \textsf{BbST}, 
while for a larger number of queries \textsf{BbST} takes the lead.
In almost all cases, our two most successful variants are several times 
faster than \textsf{ST-RMQ\textsubscript{CON}}, 
sometimes (\textsf{BbST}, relatively large $q$) reaching an order of magnitude 
gap in performance.

Table~\ref{table:partial} contains some profiling data.
Namely, cumulative percentages of the execution times for the four successive stages 
(cf.~Section~\ref{sec:offline_contraction})
of \textsf{BbST\textsubscript{CON}} with default settings, are shown.
Unsurprisingly, for a growing number of queries the relative impact of 
the sorting stage (labeled as stage 1) grows, 
otherwise the array contraction (stage 2) is dominating.
The last two stages are always of minor importance in these tests.

\begin{table}[t!]
\centering
\begin{tabular}{rrrrr}
\hline
~~$q$ (in 1000s)~~~~~~& stage 1~~~& stages 1--2~~~& stages 1--3~~~& stages 1--4~~~\\
\hline
\multicolumn{5}{c}{$n = 100\textrm{M}$}  \\
\hline
~~~10~~~~~~&  1.4~~~& 95.9~~~& 95.9~~~& 100.0~~~\\
~~320~~~~~~& 23.5~~~& 92.5~~~& 93.0~~~& 100.0~~~\\
10240~~~~~~& 65.8~~~& 88.3~~~& 89.1~~~& 100.0~~~\\
\hline
\multicolumn{5}{c}{$n = 1\textrm{G}$}  \\
\hline
~~~32~~~~~~&  0.4~~~& 99.6~~~& 99.6~~~& 100.0~~~\\
~1024~~~~~~& 13.8~~~& 96.5~~~& 96.8~~~& 100.0~~~\\
32768~~~~~~& 59.0~~~& 87.9~~~& 88.6~~~& 100.0~~~\\
\hline
\end{tabular}
\vspace{4mm}
\caption{Cumulative percentages of the execution times for the successive stages 
of \textsf{BbST\textsubscript{CON}} with the fastest serial sort (kxsort).
The default value of $k$ (512) was used.
Each row stands for a different number of queries (given in thousands).}
\label{table:partial}
\end{table}

\begin{figure}[pt]
\centerline{
\includegraphics[width=0.495\textwidth,scale=1.0]{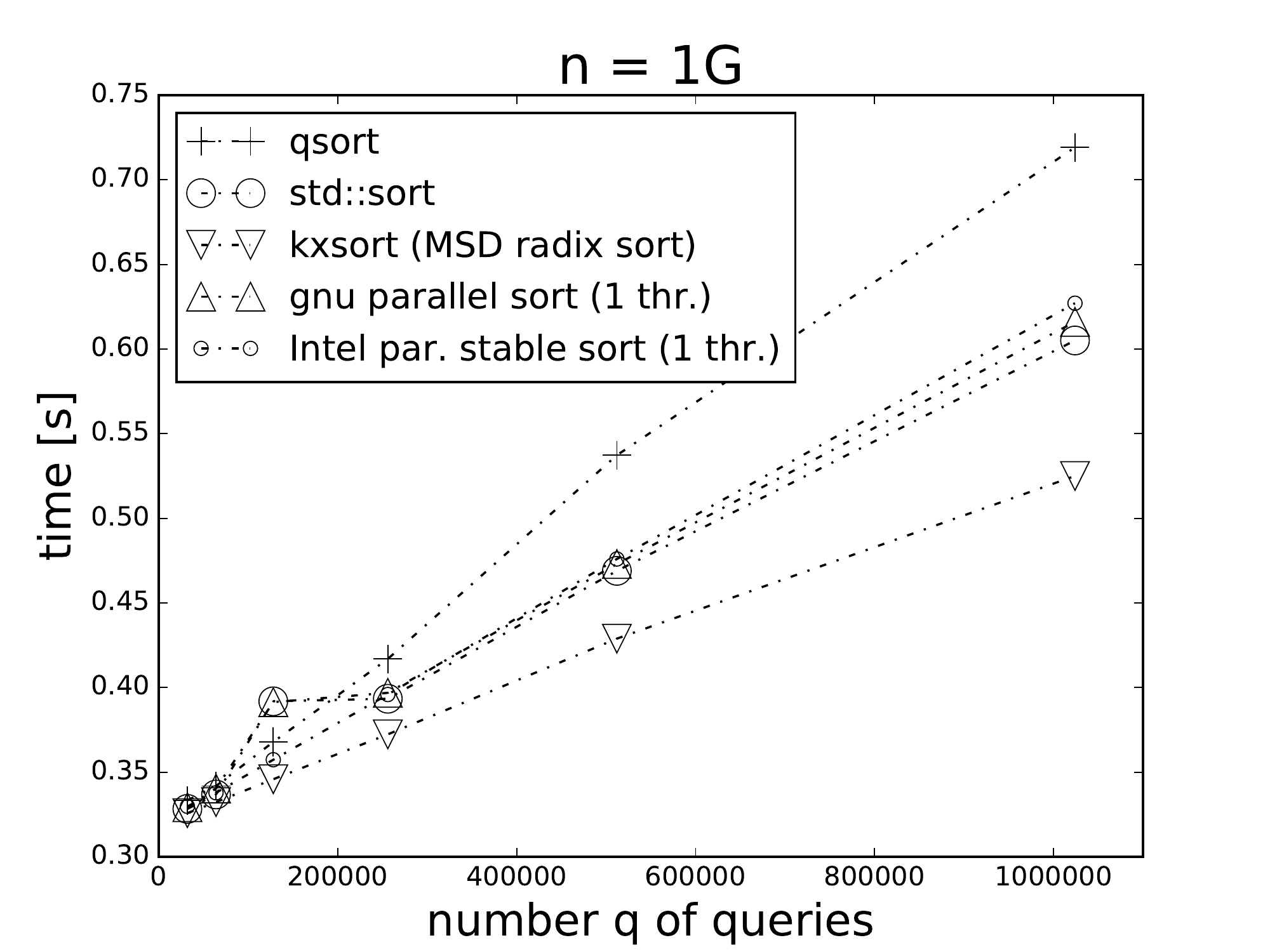}
\includegraphics[width=0.495\textwidth,scale=1.0]{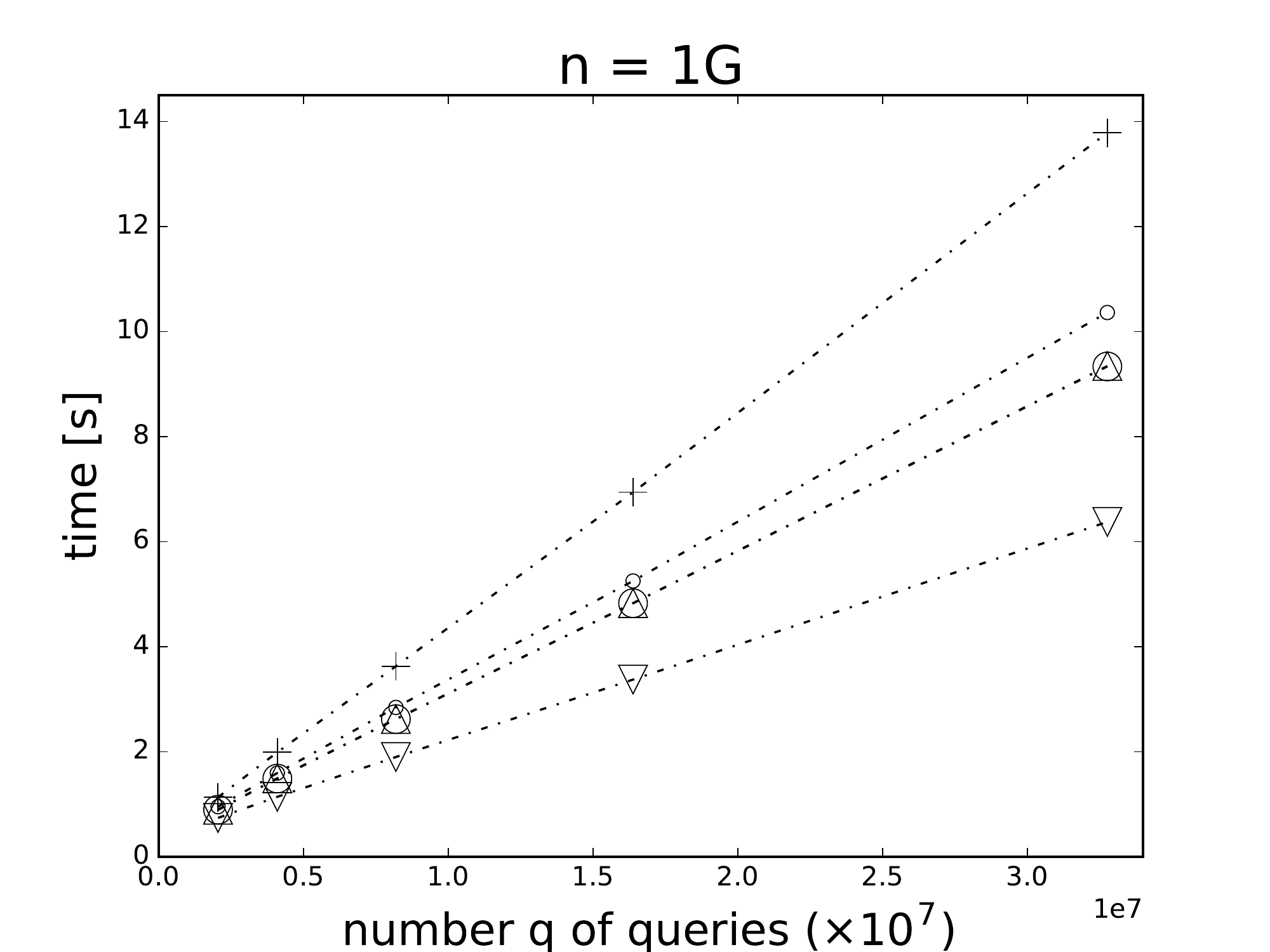}
}
\caption[Fig5]
{Impact of the sort algorithm on the running times of \textsf{BbST\textsubscript{CON}}.
The number of queries $q$ varies from $\sqrt{n}$ to $32\sqrt{n}$ (left figures) 
and from $64\sqrt{n}$ to $1024\sqrt{n}$ (right figures), 
where 
$n$ is $1\textrm{G}$.}
\label{fig:var_sort}
\end{figure}

Different sorts for \textsf{BbST\textsubscript{CON}}, in a serial regime, 
were applied in the experiment shown in Fig.~\ref{fig:var_sort}.
Namely, we tried out C++'s qsort and std::sort, kxsort, \_\_gnu\_parallel::sort 
and Intel parallel stable sort (pss).
The function qsort, as it is easy to guess, is based on quick sort.
The other sort from the C++ standard library, std::sort, implements introsort, 
which is a hybrid of quick sort and heap sort.
Its idea is to run quick sort and only if it gets into trouble on some pathological data 
(which is detected when the recursion stack exceeds some threshold), switch to heap sort.
In this way, std::sort works in $O(n\log n)$ time in the worst case.
The next contender, kxsort, is an efficient MSD radix sort.
The last two sorters are parallel algorithms, but for this test they are run with 
a single thread.
The gnu sort is a multiway mergesort (exact variant)
from the GNU libstdc++ parallel mode library.
Finally, Intel's pss is a parallel merge 
sort\footnote{\texttt{https://software.intel.com/en-us/articles/\\
a-parallel-stable-sort-using-c11-for-tbb-cilk-plus-and-openmp}}.
We use it in the OpenMP~3.0 version.

\begin{figure}[pt!]
\centerline{
\includegraphics[width=0.495\textwidth,scale=1.0]{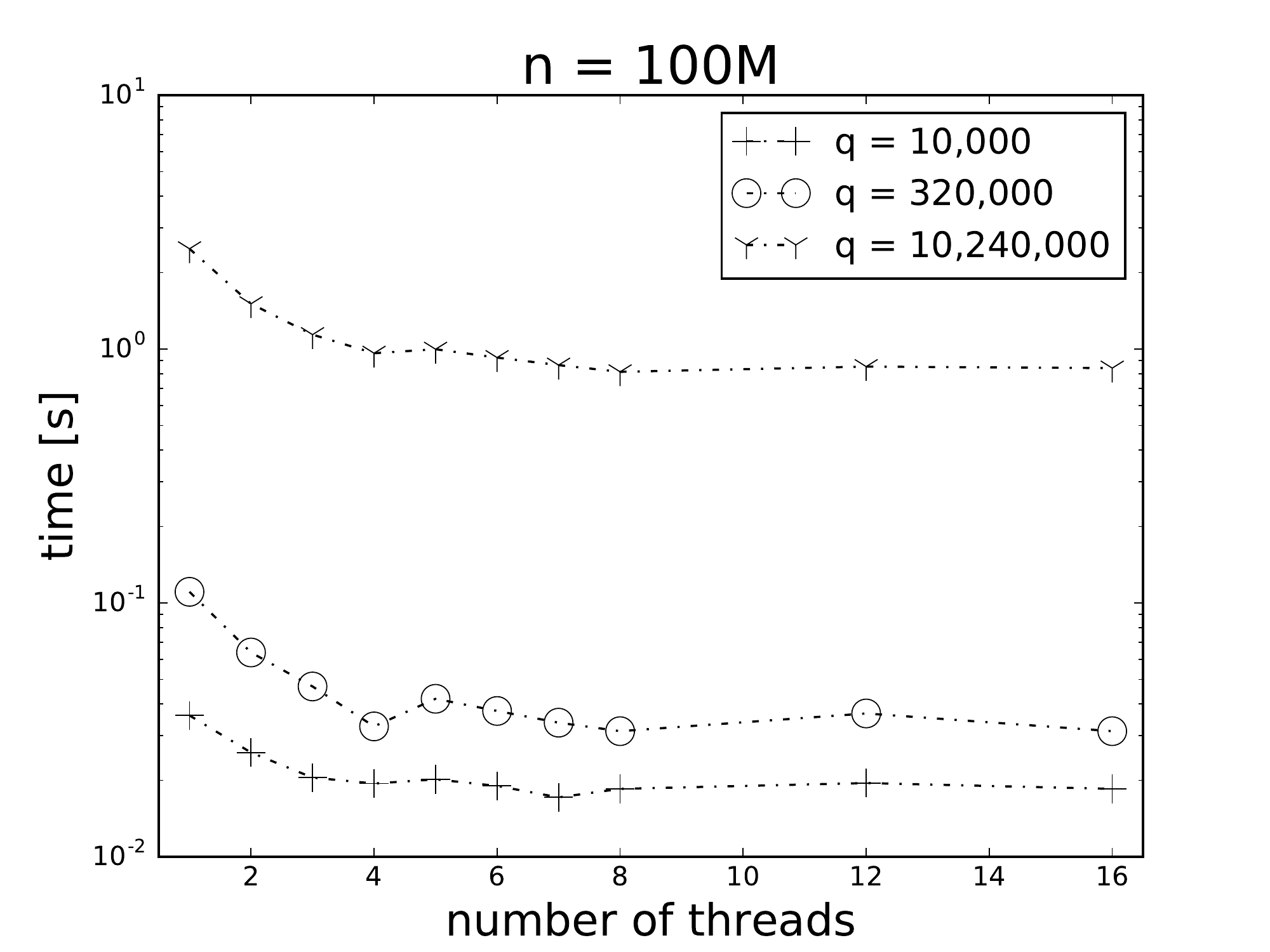}
\includegraphics[width=0.495\textwidth,scale=1.0]{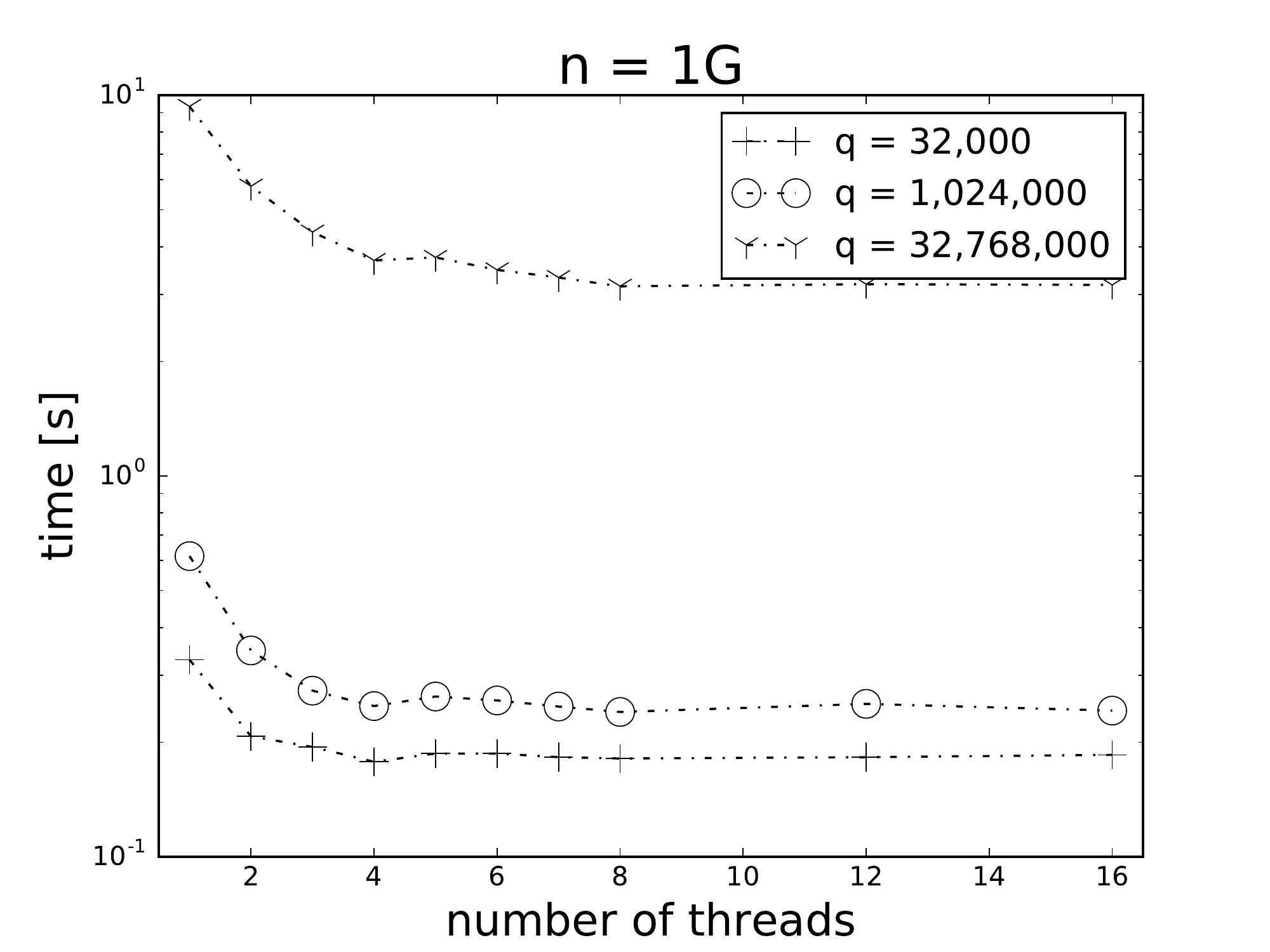}
}
\caption[Fig4]
{Impact of the number of threads in \_\_gnu\_parallel::sort 
and in creating $A_Q$ (by independent scanning for minima in 
contiguous areas of $A$)
on the overall performance of \textsf{BbST\textsubscript{CON}}, 
for different number of queries $q$, 
where $n$ is 100M (left figure) or 1G (right figure).
Note the logarithmic scale on the Y-axis.}
\label{fig:var_threads}
\end{figure}

For the last experiment with \textsf{BbST\textsubscript{CON}}, 
we ran our algorithm in a parallel mode, 
varying the number of threads in $\{1, 2, \ldots, 8, 12, 16\}$ 
(Fig~\ref{fig:var_threads}).
For sorting the queries we took the faster parallel sort, 
\_\_gnu\_parallel::sort.
The remaining stages also benefit from parallelism.
The second stage computes in parallel the minima in contiguous areas of $A$ 
and the third stage correspondingly handles blocks of $A_Q$.
Finally, answering queries is handled in an embarassingly parallel manner.

As expected, the performance improves up to 8 threads 
(as the test machine has 4 cores and 8 hardware threads), but the overall 
speedups compared to the serial variant are rather disappointing, 
around factor 2 or slightly more.

Table~\ref{table:memory} presents the memory use 
(apart from input array $A$ and the set of queries $Q$) for our variants.
\textsf{BbST} is insensitive here to $q$.
The parameter $k$ was set to 512 in the case of \textsf{BbST\textsubscript{CON}}. 
As expected, the space for \textsf{BbST\textsubscript{CON}} 
grows linearly with $q$.
For small enough $q$, \textsf{BbST\textsubscript{CON}} is more succinct 
than \textsf{BbST} (unless we run it with large $k$, which hampers the speed), 
but for the maximum tested number of queries, $q \approx 1024 \sqrt{n}$, 
\textsf{BbST} easily wins in this respect.
Finally, \textsf{BbST2} may pose a better time-space tradeoff than \textsf{BbST}.

\begin{table}[t!]
\centering
\begin{tabular}{lrr}
\hline
variant~~~~~ & \multicolumn{2}{c}{extra space as \% of the input} \\
with parameter &~~~~~$n = 100,000,000$ &~~~~~$n = 1,000,000,000$~\\
\hline
\textsf{BbST\textsubscript{CON}}, $q \approx \sqrt{n}$ & 0.10 & $ 0.03 $ \\
\textsf{BbST\textsubscript{CON}}, $q \approx 32 \sqrt{n}$ & 3.23 & 1.03 \\
\textsf{BbST\textsubscript{CON}}, $q \approx 1024 \sqrt{n}$ & 103.68 & 33.20 \\
\hline
\textsf{BbST}, $k = 512$ &  7.03 & 8.20 \\
\textsf{BbST}, $k = 1024$ & 3.32 & 3.91 \\
\textsf{BbST}, $k = 2048$ & 1.56 & 1.86 \\
\textsf{BbST}, $k = 4096$ & 0.73 & 0.88 \\
\textsf{BbST}, $k = 8192$ & 0.34 & 0.42 \\
\textsf{BbST}, $k = 16,384$ & 0.16 & 0.20 \\
\textsf{BbST}, $k = 32,768$ & 0.07 & 0.09 \\
\hline
\textsf{BbST2} $(512, 64)$ & 7.42 & 8.59 \\
\textsf{BbST2} $(4096, 256)$ & 0.83 & 0.98 \\
\textsf{BbST2} $(16384, 256)$ & 0.26 & 0.29 \\
\hline
\end{tabular}
\vspace{4mm}
\caption{Memory use for the three variants, as the percentage of the 
space occupied by the input array $A$ (which is $4n$ bytes).
The parameter $k$ was set to 512 for \textsf{BbST\textsubscript{CON}}.
}
\label{table:memory}
\end{table}

\section{Final remarks}

Computing range minimum queries over a sequence of length $n$ 
is a fundamental primitive in many compressed data structures (indexes) 
and string mining.
We proposed a very simple, yet efficient approach to this problem, 
called BbST,
adapting the well-known Sparse Table technique to work on blocks, 
with speculative block minima comparisons.
This technique alone allows to be competitive in speed to existing 
RMQ solutions, but has two drawbacks: 
it allows to obtain constant query time only in the average case 
(assuming that the range width is large enough to the block size, 
specified at construction time) 
and it requires access to the input array during the query handling.
As a next solution we thus proposed to combine our technique 
with one of the existing succinct solutions with $O(1)$ worst-case 
time queries and no access to the input array.
The resulting hybrid is still a memory frugal data structure, 
spending usually up to about $3n$ bits, and providing competitive query times, 
especially for wide ranges.
We also showed how to make our baseline data structure more compact.

Additionally, we showed how to use the BbST approach (in a standard 
or modified way) in the recently proposed scenario of offline RMQ. 
In this problem, the set of $q$ queries is given beforehand and 
if $q$ is small enough, it is not recommended to build a (heavy) 
data structure on the input array as this overhead may not be compensated.
Experimental results show that the block-based Sparse Table approach 
can be recommended also for the offline RMQ problem.
Not surprisingly, parallelization allowed to obtain extra speedups.


\section*{Acknowledgement}
The work was supported by the Polish National Science Centre under the project DEC-2013/09/B/ST6/03117 (both authors).

\bibliographystyle{psc}          
\bibliography{rmq}     

\end{document}